\documentstyle[psfig]{mn}

\title[The global structure of galactic discs]
      {The global structure of galactic discs}

\author[R.~de~Grijs]{R.~de~Grijs$^{1,2}$\\
$^1$ Kapteyn Astronomical Institute, P.O.~Box 800, 9700 AV Groningen,
The Netherlands\\
$^2$ Astronomy Dept., University of Virginia, P.O. Box 3818,
Charlottesville, VA 22903-0818, USA, grijs@virginia.edu (current
address)}

\date{Received date; accepted date}
\pubyear{1998}

\begin{document}
\maketitle

\begin{abstract} 
A statistical study of global galaxy parameters can help to improve our
understanding of galaxy formation processes.  In this paper we present
the analysis of global galaxy parameters based on optical and
near-infrared observations of a large sample of edge-on disc galaxies. \\
We found a correlation between the ratio of the radial to vertical scale
parameter and galaxy type: galaxies become systematically thinner when
going from S0's to Sc's, whereas the distribution seems to level off for
later types.  \\
The observed scale length ratios (and thus the radial colour gradients)
largely represent the galaxies' dust content.  On average the colour
gradients indicated by the scale length ratios increase from type Sa to 
at least type Sc.  For galaxy types later than Sc, the average colour
gradient seems to decrease again.  \\
The distribution of {\it K}-band (edge-on) disc central surface
brightnesses is rather flat, although with a large scatter.  However,
the latest-type sample galaxies ($T > 6$) show an indication that their
average disc central surface brightnesses may be fainter than those of
the earlier types.  This effect is probably not the result of dust
extinction.
\end{abstract}

\begin{keywords}
dust, extinction; galaxies: fundamental parameters; galaxies:
photometry; galaxies: structure
\end{keywords}

\section{A statistical analysis}

A study of the statistical properties of highly inclined, or ``edge-on''
galaxies benefits greatly from the special orientation with respect to
the line of sight of such galaxies.  Observations of edge-on galaxies
provide us with direct measurements of the luminosity and colour
distributions both perpendicular to the galaxy planes and along the
galaxies' major axes at various heights above the plane.  Indirectly,
these luminosity distributions can be related to the galaxies' density
distributions and thus their global structure.  Moreover, in-depth
knowledge of the dust distribution, and hence the optical depth of
galaxies, is important for our understanding of galaxy evolution. 

\subsection{The flattening of exponential discs}
\label{ratio.sec}

A major advantage of studying highly inclined galaxies is that one can
determine their radial and vertical scale parameters directly and
independently, since the dependence of these parameters on inclination
is smallest for the highest inclinations (e.g., van der Kruit \& Searle
1981a; de Grijs et al.  1997).  These scale parameters provide us with
information about the intrinsic shape of galaxy discs, i.e., their
flattening, in a more direct way than the canonical axis ratios. 
Moreover, since the vertical scale height, $z_0 = 2 h_z$ (where $h_z$ is
the exponential scale height), is to first order independent of radius
(e.g., van der Kruit \& Searle 1981a,b, 1982a; Kylafis \& Bahcall 1987;
Shaw \& Gilmore 1990; Barnaby \& Thronson, Jr.  1992; but see de Grijs
\& Peletier 1997), the radial to vertical scale parameter ratio,
$h_R/z_0$, can often be determined more accurately than the major to
minor axis ratio. 

By studying the scale parameter ratio statistically, we may be able to
put constraints on the disc formation processes as well as on the
stability of galaxy discs (e.g., Bottema 1993).  When considering the
physical processes that determine the scale parameters one does not
immediately expect a strong correlation between scale length and scale
height.  The scale height is likely determined by the internal, secular
evolution of the stellar velocity dispersion (e.g., van der Kruit \&
Searle 1981a; Carlberg 1987), whereas the scale length is basically the
result of the composition of the protogalaxy (Fall 1983; van der Kruit
1987). 

However, one might expect that in a larger disc, with a greater rotation
velocity, the heating of the disc stars may be more violent, thus
resulting in a larger scale height.  Therefore, one may expect a
correlation between the rotation velocity (which can be related directly
to the scale length) of a galaxy disc and the scale height, although the
precise dependence is yet unknown (see, e.g., Bottema 1993).

Thus, statistics on the ratio of scale length to scale height can be
expected to give information on the importance of the formation
processes in disc galaxies with different properties. 

Moreover, once the $h_R/z_0$ ratio is known, one may be able to
determine the (theoretical) maximum rotation of a disc from measurements
of the vertical disc dispersion (Bottema 1993).  Therefore, a
statistical treatment of the scale parameter ratio may put general
constraints on both the kinematical properties and the global stability
of galaxy discs. 

Bottema (1993) predicts that a constant value for the $h_R/z_0$ ratio
leads to a more or less constant mass-to-light ratio of the old disc,
$(M/L)_B$, under the assumption that we are dealing with exponential,
locally isothermal discs with a constant ratio of vertical to radial
velocity dispersion.  On the other hand, if we assume a linear
relationship between the old-disc absolute luminosity and the vertical
velocity dispersion, Bottema (1993) shows that, for a constant
$(M/L)_B$, the $h_R/z_0$ ratio decreases rapidly from faint galaxies to
a constant level for normal and bright galaxies.  Thus, in general, the
observed velocity dispersions imply that a constant old-disc
mass-to-light ratio results in an approximately constant scale parameter
ratio, whereas a constant scale parameter ratio also leads to a
mass-to-light ratio that is, to first order, constant. 

In fact, these predictions imply that all galaxy discs are governed by
equal mass-to-light ratios in the old stellar populations, assuming that
all galaxy discs have approximately the same colour (Bottema 1993). 

However, the assumption of a constant and equal mass-to-light ratio of
the old-disc population in disc galaxies is probably not physically
realistic, considering the range of colours observed within and among
galaxies (e.g., de Jong 1996c, and references therein).  Therefore, the
predicted relationships should be treated with caution and only be used
as general guidelines. 

\subsection{Colour gradients as diagnostics}

Broad-band colours are relatively easy to obtain and are therefore the
most widely used colour diagnostics to date.  They immediately reveal
the approximate nature of a galaxy, which is to first order determined
by its dominant stellar population and dust content.  

Although for the detailed analysis of galaxy luminosity and colour
profiles one needs to adopt {\it a priori} assumptions concerning the
evolutionary stellar population synthesis, the initial mass function,
the metallicity and the star formation history, as well as about the
dust geometry and its characteristics, de Jong (1996c) shows that the
colours formed from different broad-band combinations correlate
strongly, which indicates that these colours are probably caused by the
same physical process.  Therefore, broad-band colours can be used as
indicators of changes in the gross properties of galaxies (e.g., changes
in metallicity and/or dust contamination).  

All systematic colour differences induced by stellar population changes
and metallicity gradients are generally considerably smaller than the
reddening due to dust, however. 

\subsubsection{Radial colour gradients in edge-on disc galaxies}
\label{edgeongrad.sect}

In contrast to the large number of studies of radial colour gradients in
moderately inclined and face-on spiral galaxies (e.g., de Jong 1996c,
and references therein), the colour behaviour of highly inclined and
edge-on galaxies has not received much attention.  In highly inclined
galaxies, the study and interpretation of intrinsic colour gradients is
severely hampered by the presence of dust in the galaxy planes, which
causes the dust lane to appear as a red feature in vertical colour
profiles (e.g., Hamabe et al.  1979; Hegyi \& Gerber 1979; van der Kruit
\& Searle 1981b; Jensen \& Thuan 1982; de Grijs et al.  1997). 

In individual edge-on galaxies, it is generally found that the colours
along the major axes, i.e., the locations of the dust lanes, remain
nearly constant (e.g., Sasaki 1987; Wainscoat et al. 1990; Aoki et
al. 1991; Peletier \& Balcells 1997), although in most cases the
outermost disc regions tend to be slightly bluer on the major axis
(e.g., Sasaki 1987), which may be explained in terms of an increasingly
metal-poor population or a decreased amount of dust at larger
galactocentric distances.  Generally, as the height above the dust lane
and its embedded young disc increases, the radial colour gradients
become small or statistically insignificant (e.g., Hamabe et al. 1979,
1980; van der Kruit \& Searle 1982a,b; Jensen \& Thuan 1982; Peletier
\& Balcells 1997). 

\subsubsection{Colour gradients from scale length ratios}
\label{colgrads.sect}

Since the dust influence varies as a function of passband, scale length
ratios could be used as a diagnostic to estimate colour gradients and
the dust content of a given galaxy. 

Evans (1994) studied the effects of dust on the stellar scale length as
a function of wavelength, under the assumption that the resulting scale
length differences are solely due to dust absorption.  His models
predict that these differences are small, at least for face-on galaxies,
on the order of the observational uncertainties, and even smaller for
galaxies with a prominent bulge component.  According to his models, if
the scale height ratio between dust and stars is $\sim 0.5$ (Peletier \&
Willner 1992; Evans 1994), Evans' (1994) models exclude face-on
galaxies with $h_B/h_H \approx 2$.  On the other hand, larger ratios can
be obtained if a galaxy is inclined with respect to the line of sight. 

The measurement of blue to red scale length ratios alone will not
unambiguously reveal the dust content of a given galaxy, because any
deviation from unity can equally well be explained by an intrinsic
colour gradient, especially for face-on galaxies (Byun et al. 1994). 

\section{Sample selection, observations, and data reduction}

To study the structural parameters of edge-on spiral galaxies we   
selected a statistically complete sample taken from the Surface
Photometry Catalogue of the ESO-Uppsala Galaxies (ESO-LV; Lauberts \&
Valentijn 1989) with the following properties:
\begin{itemize}
\item their inclinations are greater than or equal to 87$^\circ$;
\item the angular blue diameters ($D_{25}$) are larger than $2.'2$;   
\item the galaxy types range from S0 to Sd, and
\item they should be non-interacting.
\end{itemize}

The inclinations were determined following Guthrie (1992), assuming a
true axial ratio $\log R_0 = 0.95$, corresponding to an intrinsic
flattening $q_0 = (b/a)_0$ of 0.11.  From this intrinsic flattening the 
inclinations {\it i} were derived by using Hubble's (1926) formula

\begin{equation}
\cos^2 i = (q^2 - q_0^2)/(1 - q_0^2),
\end{equation}
where $q = b/a$ is the observed axis ratio.

Of the total sample of 93 southern edge-ons, an arbitrary subsample of  
24 galaxies was observed in the near-infrared $K'$ band in two observing
runs of 4 and 3 nights, respectively.  The selection of these 24
observed sample galaxies depended solely on the allocation of telescope
time; the galaxies cover the southern sky rather uniformly. 

By applying a $V/V_{\rm max}$ completeness test (e.g., Davies 1990; de
Jong \& van der Kruit 1994) we derived that the ESO-LV is statistically
complete for diameter-limited samples with $D_{25}^B \ge 1.'0$.  To
check the completeness of our subsample, we calculated, based on a
limiting diameter $D_{25}^B \ge 2.'2$, that $V/V_{\rm max} = 0.502 \pm
0.253$, which implies statistical completeness. 

The near-infrared observations were obtained with the IRAC2B camera at
the ESO/MPI 2.2m telescope of the European Southern Observatory (ESO) in
Chile.  The IRAC2B camera is equipped with a Rockwell 256$\times$256
pixel NICMOS3 HgCdTe array.  For both observing runs, in July 1994 and
January 1995, we used the IRAC2B camera with Objective C, corresponding
to a pixel size of $0.''491$ (40 $\mu$m) and a field of view of
$125''\times125''$.  The near-infrared array is linear to within 1\% up
to $\sim 18,000$ counts for exposure times $> 1$ sec (see, e.g., de
Grijs 1997). 

At both runs we used the $K'$ filter available at ESO (central
wavelength $\lambda_{\rm c} = 2.15 \mu$m, bandpass $\Delta \lambda =  
0.32 \mu$m).  We chose to observe in $K'$ rather than in {\it K} band 
(with $\lambda_{\rm c} = 2.2 \mu$m, and $\Delta \lambda = 0.40
\mu$m), since the $K'$ band is almost as little affected by dust as the
{\it K} band, but has a lower sky background (Wainscoat \& Cowie 1992).

We took sky images and object frames alternately, both with equal
integration times (in sequences of 12 $\times$ 10s), and spatially   
separated by $\sim 5'$.

An overview of the near-infrared observations can be found in de Grijs
et al. (1997).

The major part of the (supplementary and additional) optical {\it B} and
{\it I}-band observations of our sample galaxies were obtained using the
Danish 1.54m telescope at ESO, equipped with a 1081$\times$1040 pixel
TEK CCD with a pixel size of 24$\mu$m (0.36 $''$/pix).  The field of
view thus obtained is $6'.5\times6.'2$.  The TEK CCD was used in slow
read-out mode in order to decrease the pixel-to-pixel noise.  The CCD
used is linear to within 1\% up to 44,000 counts and saturates at 65,536
counts. 

Gaps in the observed sample were filled in by service observations with
the Dutch 0.92m telescope at ESO, equipped with a 512$\times$512 pixel
TEK CCD.  It has a pixel size of 27 $\mu$m (0.44$''$/pix), corresponding
to a field of view of $3.'9\times3.'9$.  The CCD is linear to within 1\%
over the full 16-bit dynamic range. 

At both telescopes we used the standard Johnson {\it B} filter and the
Thuan \& Gunn (1976) {\it I} filter, which characteristics match those
of a Johnson {\it I} filter (Buser 1978).  Both telescopes were used in
direct imaging mode, at prime focus.  Details of the specific
observations can be found in Table \ref{mnras_opt.tab}. 

{
\begin{table}
\caption[ ]{\label{mnras_opt.tab}{\bf Optical observations of the sample
galaxies.} \newline Columns: (1) Galaxy name (ESO-LV); (2) Telescope
used (Dan 1.5 = Danish 1.54m; Dut 0.9 = Dutch 0.92m); (3) Date of
observation (ddmmyy) (4) Passband; (5) Exposure time (sec); (6) Seeing
FWHM $('')$.}

\begin{center}
{\tabcolsep=1.8mm
\begin{tabular}{ccccrc}
\hline
Galaxy  & Tel. & Date & Band & Exp.time & Seeing \\
(1)     & (2)  & (3)  & (4)  & (5)~~~~~ & (6)    \\
\hline 
026-G06 & Dan 1.5 & 090794 & $B$ & 2$\times$1800 & 2.1 \\
        & Dan 1.5 & 090794 & $I$ & 2$\times$900  & 1.6 \\
033-G22 & Dan 1.5 & 120194 & $B$ & 2$\times$1500 & 1.7 \\
        & Dan 1.5 & 120194 & $I$ & 2$\times$900  & 0.9 \\
041-G09 & Dan 1.5 & 100794 & $B$ & 2$\times$1800 & 1.7 \\
        & Dan 1.5 & 100794 & $I$ & 2$\times$900  & 1.4 \\
074-G15 & Dan 1.5 & 110794 & $B$ & 2$\times$1800 & 1.5 \\
        & Dan 1.5 & 110794 & $I$ & 2$\times$900  & 1.3 \\
138-G14 & Dan 1.5 & 080794 & $B$ & 2$\times$1800 & 2.2 \\
        & Dan 1.5 & 080794 & $I$ & 2$\times$900  & 1.6 \\
141-G27 & Dan 1.5 & 090794 & $B$ & 2$\times$1800 & 1.4 \\
        & Dan 1.5 & 090794 & $I$ & 2$\times$900  & 1.2 \\
142-G24 & Dan 1.5 & 110794 & $B$ & 2$\times$1800 & 1.2 \\
        & Dan 1.5 & 110794 & $I$ & 2$\times$900  & 1.4 \\
157-G18 & Dan 1.5 & 100194 & $B$ & 2$\times$1500 & 1.1 \\
        & Dan 1.5 & 100194 & $I$ & 2$\times$900  & 1.2 \\
201-G22 & Dan 1.5 & 090194 & $B$ & 2$\times$1500 & 1.3 \\
        & Dan 1.5 & 090194 & $I$ & 2$\times$900  & 1.2 \\
202-G35 & Dut 0.9 & 051094 & $B$ & 2$\times$1800 & 1.4 \\
        & Dut 0.9 & 061094 & $I$ & 2$\times$900  & 1.2 \\
235-G53 & Dan 1.5 & 110794 & $B$ & 2$\times$1800 & 1.4 \\
        & Dan 1.5 & 120794 & $I$ &  600          & 1.5 \\
        & Dut 0.9 & 170396 & $I$ & 1200          & 1.3 \\
        & Dut 0.9 & 210396 & $I$ & 1200          & 1.3 \\
240-G11 & Dut 0.9 & 051094 & $B$ & 2$\times$1800 & 1.6 \\
        & Dut 0.9 & 061094 & $B$ & 2$\times$1800 & 1.6 \\
        & Dut 0.9 & 051094 & $I$ & 2$\times$1200 & 1.6 \\
        & Dut 0.9 & 061094 & $I$ & 2$\times$1200 & 1.6 \\
263-G15 & Dan 1.5 & 120194 & $B$ & 2400          & 1.1 \\
        & Dan 1.5 & 090194 & $I$ &  900          & 1.0 \\
        & Dan 1.5 & 100194 & $I$ &  600          & 1.1 \\
        & Dan 1.5 & 120194 & $I$ &  900          & 1.0 \\
263-G18 & Dut 0.9 & 100493 & $B$ & 2400          & 1.7 \\
        & Dut 0.9 & 110493 & $R$ &  720          & 1.3 \\
        & Dut 0.9 & 190194 & $I$ & 2$\times$900  & 1.2 \\
269-G15 & Dut 0.9 & 130493 & $B$ & 2400          & 2.1 \\
        & Dut 0.9 & 130396 & $I$ & 2$\times$1200 & 1.3 \\
286-G18 & Dan 1.5 & 110794 & $B$ & 2$\times$1200 & 1.3 \\
        & Dan 1.5 & 090794 & $I$ & 2$\times$900  & 1.3 \\
288-G25 & Dut 0.9 & 051094 & $B$ & 2$\times$1800 & 1.3 \\
        & Dut 0.9 & 051094 & $I$ & 2$\times$1200 & 1.3 \\
311-G12 & Dan 1.5 & 120194 & $B$ & 2$\times$1500 & 1.0 \\
        & Dan 1.5 & 120194 & $I$ & 2$\times$900  & 1.0 \\
315-G20 & Dan 1.5 & 090194 & $B$ & 2$\times$1500 & 1.4 \\
        & Dan 1.5 & 090194 & $I$ & 2$\times$900  & 1.7 \\
321-G10 & Dut 0.9 & 280493 & $B$ & 2400          & 1.3 \\
        & Dut 0.9 & 280493 & $I$ &  300          & 1.2 \\
322-G73 & Dut 0.9 & 270493 & $B$ & 2400          & 1.6 \\
        & Dut 0.9 & 230396 & $I$ & 2$\times$1200 & 1.2 \\
322-G87 & Dut 0.9 & 270493 & $B$ & 2400          & 1.6 \\
        & Dut 0.9 & 270493 & $I$ &  300          & 1.3 \\
340-G08 & Dan 1.5 & 100794 & $B$ & 2$\times$1800 & 1.6 \\
        & Dan 1.5 & 100794 & $I$ & 2$\times$900  & 1.3 \\
340-G09 & Dan 1.5 & 100794 & $B$ & 2$\times$1800 & 2.0 \\
        & Dut 0.9 & 140396 & $I$ & 2$\times$1200 & 1.4 \\
358-G26 & Dan 1.5 & 120194 & $B$ & 2$\times$1500 & 1.0 \\
        & Dan 1.5 & 120194 & $I$ & 1200          & 1.0 \\
358-G29 & Dan 1.5 & 100194 & $B$ & 2$\times$1500 & 1.0 \\
        & Dan 1.5 & 100194 & $I$ & 2$\times$900  & 1.3 \\
\hline
\end{tabular}
}
\end{center}
\end{table}
}

\addtocounter{table}{-1}

{
\begin{table}
\caption[ ]{(Continued)}

\begin{center}
{\tabcolsep=1.8mm
\begin{tabular}{ccccrc}
\hline
Galaxy  & Tel. & Date & Band & Exp.time & Seeing \\
(1)     & (2)  & (3)  & (4)  & (5)~~~~~ & (6)    \\
\hline
377-G07 & Dut 0.9 & 120493 & $B$ & 2400          & 2.2 \\
        & Dut 0.9 & 100396 & $I$ & 2$\times$1200 & 1.2 \\
383-G05 & Dan 1.5 & 080794 & $B$ & 2$\times$1800 & 2.0 \\
        & Dan 1.5 & 080794 & $I$ & 2$\times$900  & 1.3 \\
416-G25 & Dan 1.5 & 130194 & $B$ & 2$\times$1500 & 1.2 \\
        & Dan 1.5 & 130194 & $I$ & 2$\times$900  & 1.0 \\
435-G14 & Dut 0.9 & 080493 & $B$ & 2400          & 1.6 \\
        & Dut 0.9 & 201293 & $B$ & 1000          & 1.7 \\
        & Dut 0.9 & 201293 & $I$ & 2$\times$900  & 1.0 \\
435-G25 & Dut 0.9 & 230493 & $B$ & 5400          & 1.9 \\
        & Dut 0.9 & 231293 & $B$ & 1800          & 1.8 \\
        & Dut 0.9 & 311293 & $B$ & 1800          & 1.4 \\
        & Dut 0.9 & 050194 & $B$ & 1800          & 1.5 \\
        & Dut 0.9 & 060194 & $B$ & 1800          & 1.3 \\
        & Dut 0.9 & 230493 & $I$ & 2$\times$300  & 1.7 \\
        & Dut 0.9 & 231293 & $I$ & 2$\times$900  & 1.3 \\
        & Dut 0.9 & 060194 & $I$ & 2$\times$900  & 1.2 \\
435-G50 & Dan 1.5 & 100194 & $B$ & 2$\times$1500 & 1.2 \\
        & Dan 1.5 & 100194 & $I$ & 2$\times$900  & 1.1 \\
437-G62 & Dut 0.9 & 040195 & $B$ & 2$\times$1800 & 1.3 \\
        & Dut 0.9 & 040195 & $I$ & 2$\times$1200 & 1.0 \\
444-G21 & Dut 0.9 & 180493 & $B$ & 2400          & 1.6 \\
        & Dut 0.9 & 230396 & $I$ & 2$\times$1200 & 1.2 \\
446-G18 & Dan 1.5 & 090794 & $B$ & 2$\times$1800 & 1.2 \\
        & Dan 1.5 & 090794 & $I$ & 2$\times$900  & 1.1 \\
446-G44 & Dan 1.5 & 110794 & $B$ & 2$\times$1800 & 1.1 \\
        & Dut 0.9 & 140396 & $I$ & 2$\times$1200 & 1.2 \\
460-G31 & Dan 1.5 & 080794 & $B$ &  900          & 1.2 \\
        & Dan 1.5 & 100794 & $B$ & 2700          & 1.3 \\
        & Dan 1.5 & 080794 & $I$ & 2$\times$900  & 1.5 \\
487-G02 & Dan 1.5 & 090194 & $B$ & 2$\times$1500 & 1.3 \\
        & Dan 1.5 & 090194 & $I$ & 2$\times$900  & 1.1 \\
500-G24 & Dan 1.5 & 130194 & $B$ & 2$\times$1500 & 1.2 \\
        & Dan 1.5 & 130194 & $I$ & 2$\times$900  & 1.0 \\
505-G03 & Dut 0.9 & 280493 & $B$ & 2400          & 1.3 \\
        & Dut 0.9 & 010496 & $I$ & 2$\times$1200 & 1.3 \\
506-G02 & Dut 0.9 & 280493 & $B$ & 2400          & 1.2 \\
        & Dut 0.9 & 280493 & $I$ &  300          & 1.0 \\
509-G19 & Dan 1.5 & 100794 & $B$ & 2$\times$1800 & 1.3 \\
        & Dan 1.5 & 100794 & $I$ & 2$\times$900  & 1.0 \\
531-G22 & Dut 0.9 & 270994 & $B$ & 2$\times$1800 & 1.6 \\
        & Dut 0.9 & 270994 & $I$ & 2$\times$1200 & 1.5 \\
555-G36 & Dan 1.5 & 130194 & $B$ & 2$\times$1500 & 1.2 \\
        & Dan 1.5 & 130194 & $I$ & 2$\times$900  & 1.1 \\
564-G27 & Dut 0.9 & 211293 & $B$ & 2400          & 1.4 \\
        & Dut 0.9 & 221293 & $B$ & 1800          & 1.7 \\
        & Dut 0.9 & 211293 & $I$ & 2$\times$900  & 1.4 \\
        & Dut 0.9 & 221293 & $I$ & 2$\times$900  & 1.3 \\
\hline
\end{tabular}
}
\end{center}
\end{table}
}

For all observed galaxies we determined the colour terms required for
the calibration using standard stars (Table \ref{zpo_en_zo.tab}). 

{
\begin{table}
\tabcolsep=0.5mm

\caption[ ]{\label{zpo_en_zo.tab}{\bf Calibration transformation coefficients
obtained for the different observing runs.}}
\begin{tabular}{clrl}
\hline
Pass- & \multicolumn{1}{c}{zero point} & \multicolumn{1}{c}{colour coeff.}
  & \multicolumn{1}{c}{extinction} \\
band  & \multicolumn{1}{c}{offset} &
  & \multicolumn{1}{c}{(mag/airmass)} \\
\multicolumn{1}{c}{(1)} & \multicolumn{1}{c}{(2)} & \multicolumn{1}{c}{(3)} & 
\multicolumn{1}{c}{(4)} \\
\hline
\multicolumn{4}{c}{April 1993, Dutch 0.92m telescope} \\
$B$ & ~~~20.721 $\pm$ 0.109 &   0.110 $\pm$ 0.032~~ & ~~~0.221 $\pm$ 0.030 \\
$I$ & ~~~20.606 $\pm$ 0.083 & --0.037 $\pm$ 0.015~~ & ~~~0.097 $\pm$ 0.030 \\
\hline
\multicolumn{4}{c}{December 1993/January 1994, Dutch 0.92m telescope} \\
$B$ & ~~~20.982 $\pm$ 0.087 &   0.111 $\pm$ 0.027~~ & ~~~0.242 $\pm$ 0.045 \\
$I$ & ~~~20.853 $\pm$ 0.048 & --0.036 $\pm$ 0.013~~ & ~~~0.104 $\pm$ 0.029 \\
\hline
\multicolumn{4}{c}{January 9--13, 1994, Danish 1.54m telescope} \\
$B$ & ~~~22.331 $\pm$ 0.040 &   0.124 $\pm$ 0.031~~ & ~~~0.105 $\pm$ 0.030 \\
$I$ & ~~~22.351 $\pm$ 0.040 & --0.035 $\pm$ 0.022~~ & ~~~0.021 $\pm$ 0.013 \\
\hline
\multicolumn{4}{c}{July 9--13, 1994, Danish 1.54m telescope} \\
$B$ & ~~~22.984 $\pm$ 0.056 &   0.157 $\pm$ 0.006~~ & ~~~0.308 $\pm$ 0.051 \\
$I$ & ~~~22.497 $\pm$ 0.028 & --0.042 $\pm$ 0.018~~ & ~~~0.030 $\pm$ 0.022 \\
\hline
\multicolumn{4}{c}{September/October 1994, Dutch 0.92m telescope} \\
$B$ & ~~~21.232 $\pm$ 0.059 &   0.058 $\pm$ 0.027~~ & ~~~0.210 $\pm$ 0.042 \\
$I$ & ~~~20.972 $\pm$ 0.048 & --0.060 $\pm$ 0.018~~ & ~~~0.064 $\pm$ 0.023 \\
\hline
\multicolumn{4}{c}{January 1995, Dutch 0.92m telescope} \\
$B$ & ~~~20.900 $\pm$ 0.043 &   0.144 $\pm$ 0.030~~ & ~~~0.242 $\pm$ 0.449 \\
$I$ & ~~~20.896 $\pm$ 0.017 & --0.025 $\pm$ 0.015~~ & ~~~0.065 $\pm$ 0.023 \\
\hline
\multicolumn{4}{c}{March 1996, Dutch 0.92m telescope} \\
$I$ & ~~~20.928 $\pm$ 0.031 & --0.037 $\pm$ 0.018~~ & ~~~0.093 $\pm$ 0.032 \\
\hline
\hline
\multicolumn{4}{c}{July 9--12, 1994, ESO/MPI 2.2m telescope} \\
$K$ & ~~~20.772 $\pm$ 0.088 & & ~~~0.002 $\pm$ 0.035  \\
\hline
\multicolumn{4}{c}{February 24--26, 1995, ESO/MPI 2.2m telescope} \\
$K$ & ~~~20.697 $\pm$ 0.080 & & ~~~0.003 $\pm$ 0.029 \\
\hline
\end{tabular}
\\
{\sc Note:} \\
$^1)$ Although the errors in the zero-point offsets quoted may be
considerable, the errors obtained on the individual nights were
significantly smaller. Therefore, the latter were used for the data
reduction. \\
\end{table}
}

During the reduction of the near-infrared observations, each sky frame
was compared with the two sky frames taken nearest in time in order to
detect stars in the sky frames.  These stars were filtered out by using
a median filter and thus the resulting cleaned sky images are very
similar to the actual sky contributions.

To circumvent the effects of bad pixels and to obtain accurate
flatfielding we moved the object across the array between subsequent
exposures. Therefore, for most galaxies mosaicing of either 4 or 8
image frames was required to obtain complete galaxy images. The   
mosaicing was done by using common stars in the frames to determine the
exact spatial offsets. In the rare case that no common stars could be
determined, we used the telescope offsets as our mosaicing offsets. The
overlapping area was used to determine the adjustment of sky levels
needed by means of a least squares fit.

Bad pixels and bad areas on the array were masked out and not considered
during the entire reduction process.  Only after mosaicing was finished,
the areas that still did not contain any valid observations were
interpolated by a 2-dimensional linear plane fit (see Peletier [1993]
for a detailed description of the reduction method used).

The calibration of the near-IR observations was done by using the
SAAO/ESO/ISO Faint Standard Stars (Carter \& Meadows 1995).  We used
the corrections published by Wainscoat \& Cowie (1992) to transform the
$K'$ measurements to the {\it K} band.  The accuracy of the $K'$-band   
zero-point offsets we could reach was $\sim 0.08$ mag at both observing
runs. The limiting factors here were flatfielding errors.

The optical images were reduced following standard reduction procedures
(see de Grijs \& van der Kruit 1996); for the calibration of these
observations Landolt fields were used (Landolt 1992).  The optical
calibration could be done to an accuracy of $\sim 0.03 - 0.05$ mag,
depending on the telescope and observing run. 

Both our optical observations and the $K'$-band data were taken at
photometric (parts of) nights. 

\subsection{Internal and external consistency checks}

\subsubsection{Internal consistency checks}

\begin{figure*}
\psfig{figure=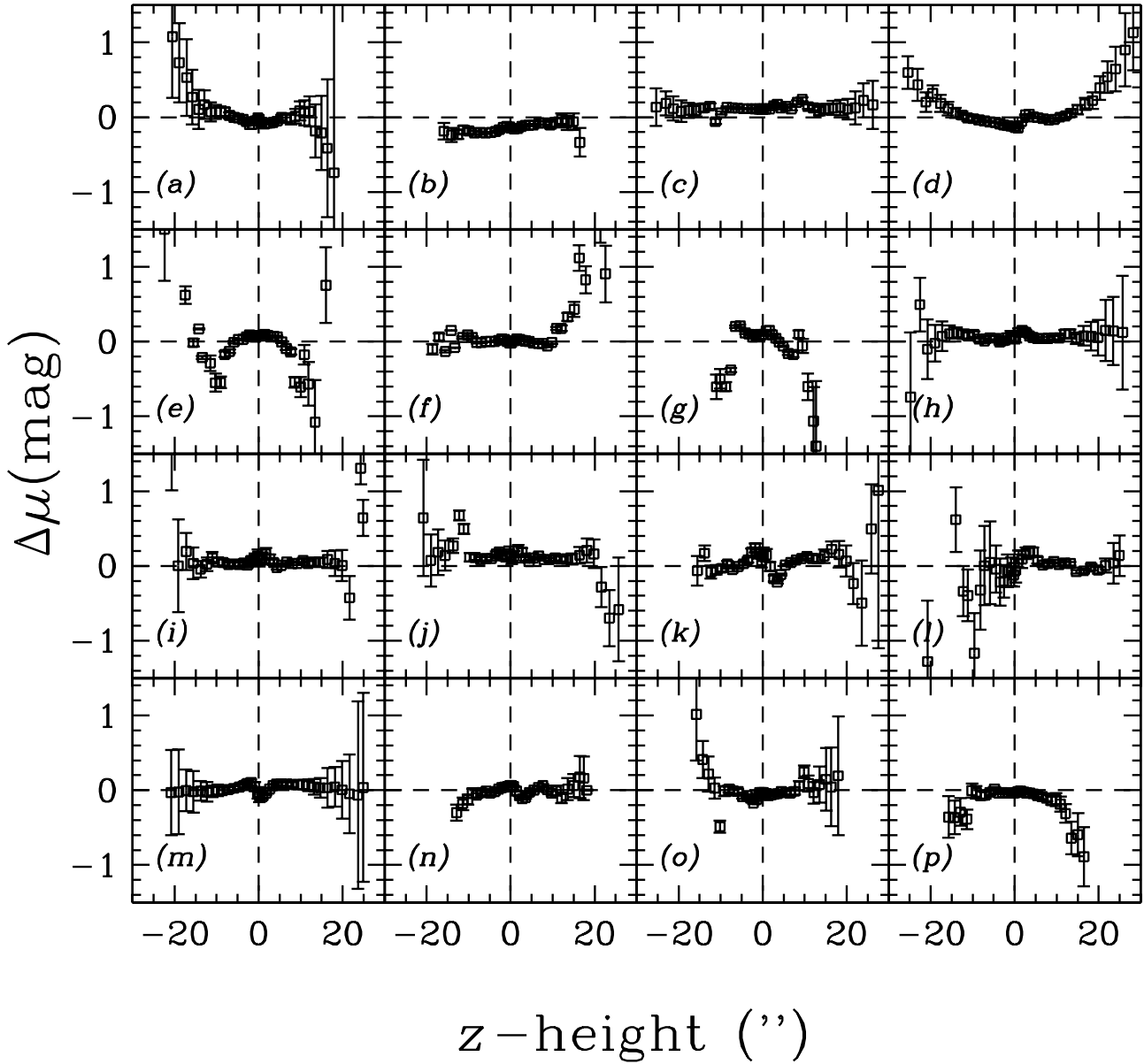,width=18cm}
\vspace*{-6.8cm}
\caption[]{\label{optcomp.fig}Comparison of the optical minor axis
profiles obtained on different nights or with different telescopes.
Details of the observations are given in Table  \ref{optcomp.tab}}
\end{figure*}

To get an impression of the quality of our observations, we performed  
both internal and external consistency checks.  We observed some of our
sample galaxies more than once through the same filter either with the  
same telescope but on different nights, or with a different telescope 
during an other observing run.  These observations provide us with an
internal consistency check.  Fig.  \ref{optcomp.fig} shows the
difference between minor axis profiles observed in optical passbands on
different nights or with a different telescope, with independent
determinations of the calibration coefficients.  The error estimates are
based on the sky noise.  Details of the specific observations that were
used to determine the difference profiles of Fig.  \ref{optcomp.fig} are
given in Table \ref{optcomp.tab}.

{
\begin{table}
\caption[ ]{\label{optcomp.tab}{\bf Details of the observations used for the
internal consistency check of Fig. \ref{optcomp.fig}}.  
\newline Columns: (1) Panel of Fig.  \ref{optcomp.fig}; (2) Galaxy name
(ESO-LV); (3) Passband; (4) and (5) Telescope used (Dan = Danish 1.54m;
Dut = Dutch 0.92m) and date of observation (ddmmyy)}

\begin{center}
{\tabcolsep=1.8mm
\begin{tabular}{ccccc}
\hline
Figure & Galaxy & Band & \multicolumn{2}{c}{Telescope/date} \\
       & (ESO)  &      & (Profile 1)  & (Profile 2)         \\
(1)    & (2)    & (3)  & (4)          & (5)                 \\
\hline 
{\it (a)} & 235-G53 & {\it I} & Dan 120794 & Dut 170396 \\
{\it (b)} & 235-G53 & {\it I} & Dut 170396 & Dut 210396 \\
{\it (c)} & 240-G11 & {\it B} & Dut 051094 & Dut 061094 \\
{\it (d)} & 240-G11 & {\it I} & Dut 051094 & Dut 061094 \\
{\it (e)} & 263-G15 & {\it I} & Dan 100194 & Dan 110194 \\
{\it (f)} & 263-G15 & {\it I} & Dan 100194 & Dan 130194 \\
{\it (g)} & 435-G14 & {\it B} & Dut 080493 & Dut 201293 \\
{\it (h)} & 435-G25 & {\it B} & Dut 230493 & Dut 231293 \\
{\it (i)} & 435-G25 & {\it B} & Dut 311293 & Dut 050194 \\
{\it (j)} & 435-G25 & {\it B} & Dut 311293 & Dut 060194 \\
{\it (k)} & 435-G25 & {\it B} & Dut 311293 & Dut 230493 \\
{\it (l)} & 435-G25 & {\it I} & Dut 230493 & Dut 231293 \\
{\it (m)} & 435-G25 & {\it I} & Dut 230493 & Dut 060194 \\
{\it (n)} & 460-G31 & {\it B} & Dan 090794 & Dan 110794 \\
{\it (o)} & 564-G27 & {\it B} & Dut 211293 & Dut 221293 \\
{\it (p)} & 564-G27 & {\it I} & Dut 211293 & Dut 221293 \\
\hline
\end{tabular}
}
\end{center}
\end{table}
}

\begin{figure}
\psfig{figure=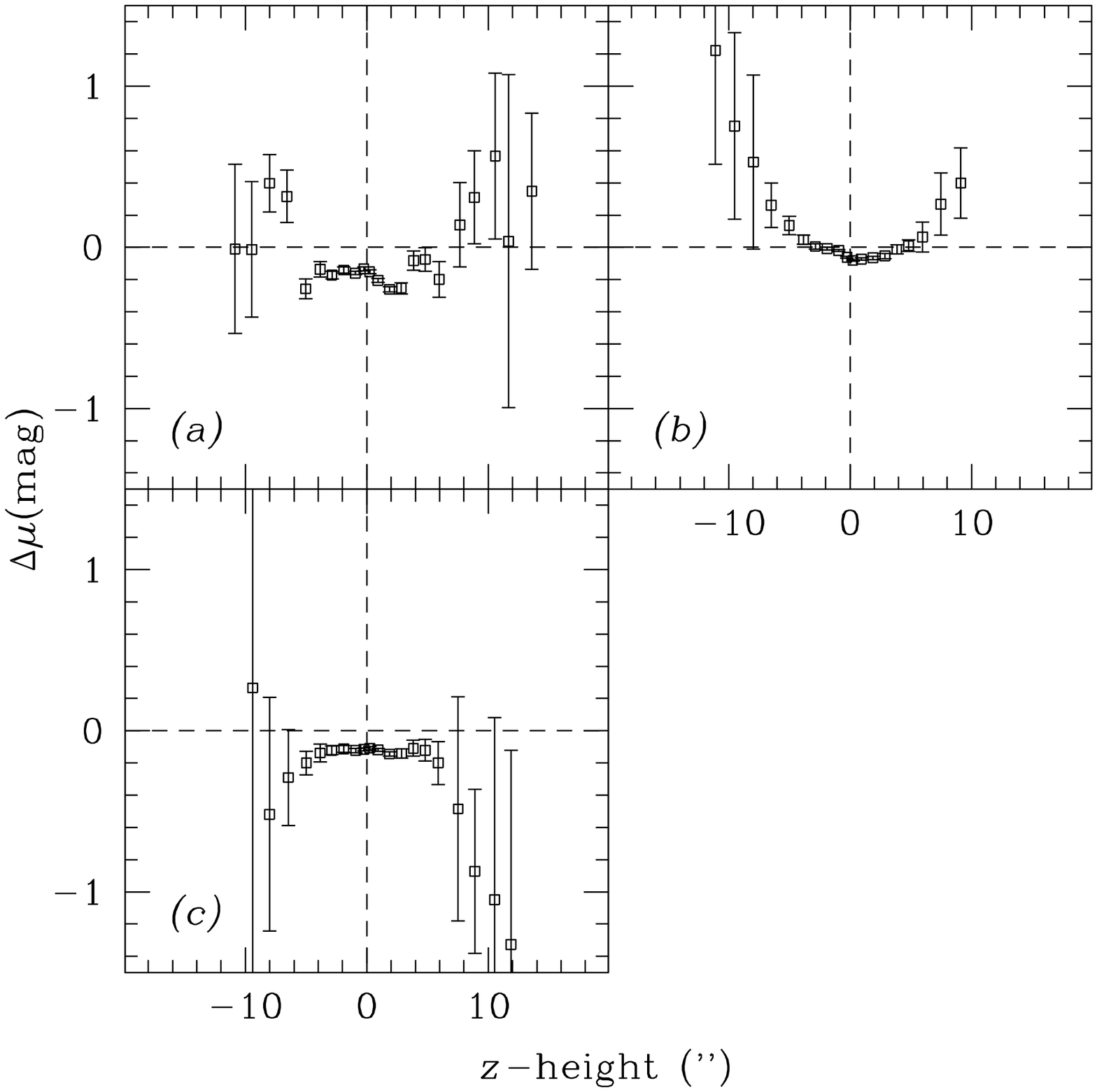,width=8.5cm}
\caption[]{\label{nircomp.fig}Comparison of near-infrared minor axis
profiles obtained on different nights or with different telescopes. 
{\it (a)} $K'$-band observations of ESO 263-G15, observed on February 24
and 25, 1995; {\it (b)} Observations of ESO 286-G18 in $K'$, observed on
July 9 and 11, 1995; {\it (c)} $K'$-band observations of ESO 446-G18,
obtained on July 9 and 12, 1995.}
\end{figure}

In Fig.  \ref{nircomp.fig} a comparison between near-infrared
observations obtained on different nights is shown.  We notice a
significant difference between the individual observations of both ESO
263-G15 and ESO 446-G18 of up to 0.2 mag.  This difference is likely due
to the slightly varying response of the IRAC2B detector as a function of
position across the array.  The accuracy of the final, combined $K'$
images is within the observational zero-point uncertainties, however.

The difference minor axis profiles in Figs.  \ref{optcomp.fig} and   
\ref{nircomp.fig} show that in general the internal consistency is well
within the zero-point and sky errors; the latter ones being generally of
order 0.1--0.2 mag.  In some cases differences in seeing show up in the
central parts of the minor axis profiles as pronounced features in the  
difference profiles.  We conclude that, although our observations are  
internally consistent, care has to be taken when interpreting the light 
at large distances from the galaxy planes, since at those distances the
difference between individual observations can amount up to a few tenths
of a magnitude.  In a few cases, like for ESO 315-G20 and ESO 435-G14,
the errors are large due to the low signal-to-noise ratio in either of
the minor axis profiles used.

\subsubsection{Comparison with published observations}

Luminosity profiles of edge-on galaxies observed with modern detectors
are scarcely available in the recent literature.  For the galaxies in
our present sample, a few individual measurements of optical luminosity
profiles, either radially, vertically, or azimuthally averaged are 
available, although they are not sufficiently well documented to prove
useful for a detailed comparison to our photometry.  Fortunately,
however, Mathewson et al.  (1992) and Mathewson \& Ford (1996), as well
as Barteldrees \& Dettmar (1994) have published detailed photometry of a
significant fraction of our sample galaxies.  From the detailed
comparison of our photometry to theirs, as discussed in the following 
sections, our main conclusion is that, within the observational errors,
we can reproduce the results published in the literature.

In de Grijs et al.  (1997) we compared our observations of the large
southern edge-on galaxy ESO 435-G25 in the near-infrared {\it K} band
with those of Wainscoat et al.  (1989).  Although our observations of
ESO 435-G25 are of a much higher quality and were taken with a much
higher resolution, we find remarkably good agreement between Wainscoat
et al.'s (1989) and our {\it K}-band observations. 

\paragraph{Comparison with Mathewson et al. (1992)}

A large and homogeneous set of observations of southern Sb--Sc galaxies
has been published recently by Mathewson et al.  (1992) and Mathewson \&
Ford (1996), of which we used their azimuthally averaged luminosity
profiles for a detailed comparison with our observations, see Fig.  3 of
de Grijs et al.  (1997).  Unfortunately, since Mathewson et al.  (1922)
and Mathewson \& Ford (1996) did not tabulate the ellipticities nor the
position angles used for the individual ellipses obtained for each
galaxy, we can at best compare azimuthally averaged profiles which were
obtained with the same free parameters. 

In general, we find that the differences between our and Mathewson's
measurements are small, although clear deviations are appreciated in a
number of cases.  In particular for those galaxies for which the
difference between our and Mathewson's profiles is relatively large, we
used the individual observations to check our results. 

\paragraph{Comparison with Barteldrees \& Dettmar (1994)}

Barteldrees \& Dettmar (1994) presented detailed optical surface
photometry for a sample of 27 edge-on disc galaxies, which has 8
galaxies in common with our sample. The observations were done in the  
Thuan \& Gunn (1976) {\it g} and {\it r} bands, for which the
transformation to the Johnson {\it B} band is obtained as follows
(Barteldrees \& Dettmar 1994; Thuan \& Gunn 1976):
\begin{equation}
g = B - 0.423 (B - R) - 0.14
\end{equation}
and
\begin{equation}
r = B - 1.78  (B - V) + 0.37
\end{equation}

\begin{figure*}
\hspace*{1.3cm}
\psfig{figure=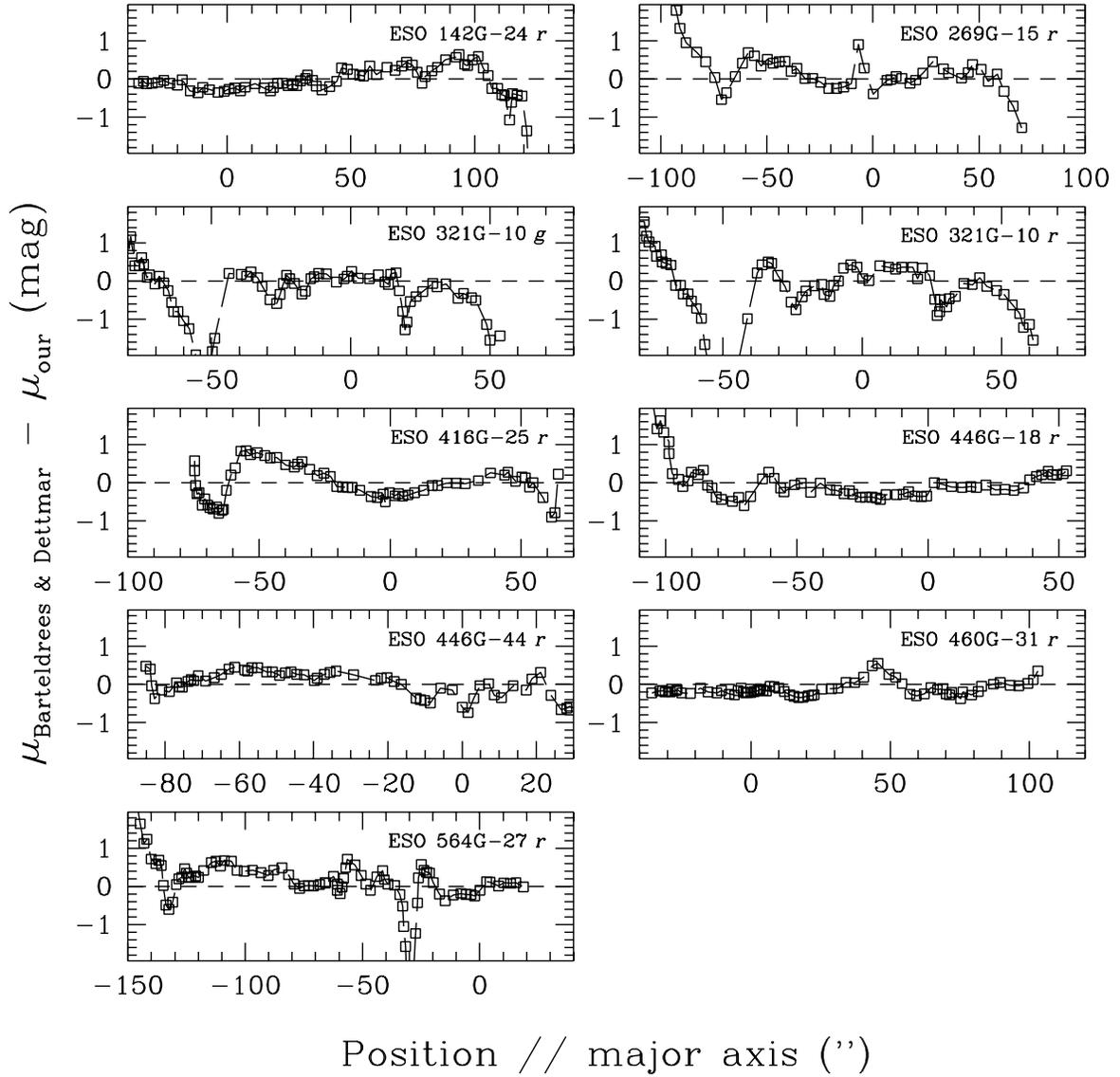,width=16cm}
\caption[]{\label{vgldettmar.fig}Comparison between major axis {\it r} 
(and {\it g})-band profiles published by Barteldrees \& Dettmar (1994), 
and those extracted from our observations.  The width of our major axis 
profiles was chosen to be in the order of the seeing FWHM, since
Barteldrees \& Dettmar (1994) did not specify this.}
\end{figure*}

We extracted surface brightness profiles along the galaxies' major axes,
and compared these to the major axis profiles plotted by Barteldrees \&
Dettmar (1994).  To do so, we calculated the transformations from the
standard Johnson {\it B} band to Thuan \& Gunn (1976) {\it g} and {\it
r} band, using {\it B--V} and {\it B--R} colours taken from the ESO-LV. 
The resulting difference profiles are shown in Fig. 
\ref{vgldettmar.fig}.  The close agreement resulting from this
comparison leads us to the conclusion that our photometry reproduces
that of Barteldrees \& Dettmar (1994) to within the observational
uncertainties. 

\section{Approach}
\label{approach5.sec}

\subsection{Determination of the exponential scale length}
\label{hlength.sect}

Knapen \& van der Kruit (1991), among others, have shown that the scale
lengths of a particular galaxy, determined by different authors, may
vary by $\approx 20$\%.  The scale length determinations depend heavily
on the radial fitting range, which is due to the fact that in general
the radial profiles are not exactly exponential.  Moreover, a patchy
dust distribution in a particular galaxy can easily lead to varying
estimates of the scale length, depending on the way the data is reduced
(e.g., Giovanelli et al. 1994). 

Since we wish to compare scale lengths determined in different passbands
within and among galaxies, we need to choose the radial regions over
which the disc exponential scale lengths are fitted in a consistent
way. 

The method that is generally used to determine the scale lengths of disc
galaxies is by applying a bulge-disc decomposition algorithm to the
azimuthally (elliptically) averaged radial luminosity profiles. 
However, when dealing with highly inclined or edge-on galaxies, the
results from this technique start to be affected by the vertical as well
as the radial luminosity components, which are characterised by
significantly different scale parameters.  Since the vertical scale
height in a typical disc galaxy is generally an order of magnitude
smaller than the radial scale length (see Sect.  \ref{scalepars.sect}),
the scale lengths derived from elliptically averaged luminosity profiles
will be smaller than the true scale lengths. 

Also, due to line-of-sight integration, the projection of a radially
exponential disc will be a function of the form 
\begin{equation}
\label{projected.eq}
L(R) = L_0 (R/h_R) K_1(R/h_R)
\end{equation}
(van der Kruit \& Searle 1981a), where $L_0$ is the central surface
luminosity, {\it R} the galactocentric distance, $h_R$ the radial scale
length, and $K_1$ the modified Bessel function of the first kind, which
simplifies to
\begin{eqnarray}
(R/h_R) K_1(R/h_R) &=& 1 + {R^2 \over 2 h_R^2} \ln(R / 2h_R), \nonumber \\
&~& \qquad \qquad (R/h_R \ll 1); \\
(R/h_R) K_1(R/h_R) &=& \Bigl({\pi R \over 2 h_R}\Bigr)^{1/2} 
\exp(-R/h_R), \nonumber \\
&~& \qquad \qquad (R/h_R \gg 1).
\end{eqnarray}
in the limits.  This means that the projected radial luminosity profile
will be flattened towards the galaxy centre.  Therefore, unless the
projected scale lengths are obtained from the outer parts of the discs
only, where $R/h_R \gg 1$, the resulting scale lengths will be estimated
too large. 

If a strong dust lane is present, the ellipse fitting is severely
hampered by extinction effects that are hard to deal with numerically. 
It is generally assumed that the dust in a galaxy is, just like the
stars, distributed in a disc with an exponentially decreasing density as
a function of radius, although with a smaller scale height than that of
the stellar component (e.g., Disney et al.  1989).  Physically, this
means that extinction plays a more prominent role near the galaxy centre
than in the outer parts, thus resulting in an observed stellar scale
length that is larger than the intrinsic scale length. 

Although these effects counteract each other, the net result for a given
galaxy depends on both the scale parameter ratio and the amount of dust,
which both correlate with galaxy type (see, e.g., Sect. 
\ref{scalepars.sect} and de Grijs et al.  1997).  However, since the
effects are not yet physically well understood, one should try to avoid
using the scale lengths obtained from elliptically averaged luminosity
profiles. 

Alternatively, one can determine the scale lengths from luminosity
profiles extracted at some distance from the galaxy planes and parallel
to the galaxies' major axes.  The main advantage of this method is that
these profiles are less affected by dust than the elliptically averaged
profiles.  Moreover, if we assume that the disc thickness is constant to
first order, as was argued in Sect.  \ref{ratio.sec} (see also de Grijs
\& van der Kruit 1996; de Grijs \& Peletier 1997), the intrinsic scale
length obtained from the old-disc population at some distance from the
plane will be essentially the same as the one at lower {\it z}
distances.  In the plane the situation becomes more complex due to the
presence of a young stellar population.  However, the young population
is generally confined to the region very close to the plane, and likely
has a scale height of only $\sim 100$pc (e.g., Hamabe \& Wakamatsu
1989). 

\subsection{Determination of the exponential scale height}

In the analysis of the vertical light distribution we distinguished
between the different sides of the galaxy planes, to avoid possible dust
contamination in the case of not perfectly edge-on orientations.  In de
Grijs \& van der Kruit (1996) the basic reduction process was described
in detail. 

In de Grijs et al.  (1997) we show that the vertical profiles are more
peaked than expected for an isothermal distribution, and only marginally
less peaked than exponential.  The analysis shows that we can safely
approximate the profile's shape with an exponential distribution at {\it
z} heights greater than 1.5 scale heights.  We fitted the vertical
profiles out to 4 scale heights, thereby taking into account the
possible presence of underlying thick disc components.  Comparison for
24 galaxies with the {\it I--K} colour images in de Grijs et al.  (1997)
shows that in this vertical region the contamination of the stellar
light by dust extinction can also be considered to be negligible. 

Furthermore, in this vertical region the results obtained from the {\it
B} and the {\it I}-band data do not differ significantly (de Grijs \&
Peletier 1997).  Thus, in this region both the red and the blue light
is dominated by the old-disc population, and the vertical luminosity
profiles are not affected by, e.g., dust or the contribution of a young
stellar population. 

\section{Results}

{
\begin{table*}

\vspace*{-0.5cm}

\caption[ ]{\label{global1.tab}{\bf Scale parameters of the sample
galaxies}\\
Columns: (1) Galaxy name (ESO-LV); (2) and (3) {\it I}-band exponential
scale height (the errors are of order $0.''03$); (4)--(7) {\it B}-band
exponential scale length and measurement error; (8)--(11) {\it I}-band
exponential scale length and measurement error; (12)--(15) $K'$-band
exponential scale length and measurement error; (16); Maximum rotational
velocity (Mathewson et al., 1992).\\ {\sl All values were obtained using
a radial fitting range between 1 and 4 {\it K}-band scale lengths,
$h_{R,K}$ ({\it I}-band if no {\it K}-band data was available).}}

\begin{center}
\tabcolsep=1mm

\begin{tabular}{lrcrrrrcrrrrcrrrrc}
\hline
\multicolumn{1}{c}{Galaxy} & & & \multicolumn{4}{c}{{\it B} band} & &
\multicolumn{4}{c}{{\it I} band} & & \multicolumn{4}{c}{{\it K} band} \\
\cline{4-7} \cline{9-12} \cline{14-17}
\multicolumn{1}{c}{name} & \multicolumn{2}{c}{$h_{z,I}$} &
\multicolumn{1}{c}{$h_{R,B}$} & \multicolumn{1}{c}{$\pm$} &
\multicolumn{1}{c}{$h_{R,B}$} & \multicolumn{1}{c}{$\pm$} & &
\multicolumn{1}{c}{$h_{R,I}$} & \multicolumn{1}{c}{$\pm$} &
\multicolumn{1}{c}{$h_{R,I}$} & \multicolumn{1}{c}{$\pm$} & &
\multicolumn{1}{c}{$h_{R,K}$} & \multicolumn{1}{c}{$\pm$} &
\multicolumn{1}{c}{$h_{R,K}$} & \multicolumn{1}{c}{$\pm$} &
\multicolumn{1}{c}{$v_{\rm rot,max}$} \\
\cline{2-3}
\noalign{\vspace{2pt}}
\multicolumn{1}{c}{(ESO)} & \multicolumn{1}{c}{$('')$} &
\multicolumn{1}{c}{($h^{-1}$ kpc)} & \multicolumn{2}{c}{$('')$} &
\multicolumn{2}{c}{($h^{-1}$ kpc)} & & \multicolumn{2}{c}{$('')$} &
\multicolumn{2}{c}{($h^{-1}$ kpc)} & & \multicolumn{2}{c}{$('')$} &
\multicolumn{2}{c}{($h^{-1}$ kpc)} & \multicolumn{1}{c}{(km s$^{-1}$)} \\
\multicolumn{1}{c}{(1)} & \multicolumn{1}{c}{(2)} &
\multicolumn{1}{c}{(3)} & \multicolumn{1}{c}{(4)} &
\multicolumn{1}{c}{(5)} & \multicolumn{1}{c}{(6)} &
\multicolumn{1}{c}{(7)} & & \multicolumn{1}{c}{(8)} &
\multicolumn{1}{c}{(9)} & \multicolumn{1}{c}{(10)} &
\multicolumn{1}{c}{(11)} & & \multicolumn{1}{c}{(12)} &
\multicolumn{1}{c}{(13)} & \multicolumn{1}{c}{(14)} &
\multicolumn{1}{c}{(15)} & \multicolumn{1}{c}{(16)} \\
\hline 
026-G06 &  3.40 & 0.38 &  31.84 &  4.70 &  3.58 & 0.53 & & 26.21 &  2.94 &  2.95 & 0.33 & & 18.69 &  2.05 &  2.10 &  0.23 &  100 \\
033-G22$^1$ &  2.45 & 0.47 &  40.52 &  3.42 &  7.72 & 0.65 & & 22.12 &  1.23 &  4.21 & 0.23 & &   --- &   --- &   --- &   --- &  113 \\
041-G09 &  4.42 & 0.76 &  42.62 &  2.74 &  7.36 & 0.47 & & 35.16 &  1.80 &  6.07 & 0.31 & & 37.75 &  2.10 &  6.52 &  0.36 &  182 \\
074-G15$^2$ & 10.84 & 0.22 & 231.68 & 48.82 &  4.83 & 1.02 & &173.23 & 27.92 &  3.61 & 0.58 & &   --- &   --- &   --- &   --- &  --- \\
138-G14$^2$ &  7.21 & 1.16 & 107.67 & 22.44 & 17.33 & 3.61 & & 60.91 &  6.91 &  9.80 & 1.11 & & 42.12 & 14.36 &  6.78 &  2.31 &  106 \\
141-G27 &  4.16 & 0.29 &  49.87 &  5.00 &  3.51 & 0.35 & & 37.42 &  3.01 &  2.63 & 0.21 & & 25.98 &  4.87 &  1.83 &  0.34 & ~~87 \\
142-G24 &  5.78 & 0.57 &  45.44 &  1.26 &  4.47 & 0.12 & & 35.49 &  0.79 &  3.49 & 0.08 & & 29.75 &  1.15 &  2.93 &  0.11 &  121 \\
157-G18 &  5.01 & 0.33 &  49.22 &  2.29 &  3.20 & 0.15 & & 38.53 &  0.99 &  2.50 & 0.06 & & 36.19 &  1.78 &  2.35 &  0.12 & ~~91 \\
201-G22 &  3.01 & 0.65 &  31.08 &  0.81 &  6.67 & 0.17 & & 21.98 &  0.42 &  4.72 & 0.09 & & 19.62 &  0.82 &  4.21 &  0.18 &  165 \\
202-G35 &  3.44 & 0.26 &  24.40 &  2.44 &  1.81 & 0.18 & & 21.22 &  1.08 &  1.57 & 0.08 & &   --- &   --- &   --- &   --- &  116 \\
235-G53 &  6.07 & 1.49 &  29.92 &  8.12 &  7.32 & 1.99 & & 23.45 &  3.64 &  5.74 & 0.89 & &   --- &   --- &   --- &   --- &  --- \\
240-G11 &  3.77 & 0.62 &  67.87 &  4.24 & 11.15 & 0.70 & & 41.34 &  1.57 &  6.79 & 0.26 & &   --- &   --- &   --- &   --- &  241 \\
263-G15 &  3.49 & 0.37 &  60.35 &  6.40 &  6.47 & 0.69 & & 36.09 &  2.03 &  3.87 & 0.22 & & 32.24 &  2.15 &  3.45 &  0.23 &  --- \\
263-G18 &  2.95 & 0.56 &  32.85 &  2.56 &  6.21 & 0.48 & & 24.27 &  1.59 &  4.59 & 0.30 & &   --- &   --- &   --- &   --- &  --- \\
269-G15 &  3.30 & 0.54 &  33.30 &  1.16 &  5.42 & 0.19 & & 24.96 &  0.82 &  4.07 & 0.13 & &   --- &   --- &   --- &   --- &  176 \\
286-G18 &  3.65 & 1.67 &  20.19 &  0.76 &  9.21 & 0.35 & & 17.55 &  0.46 &  8.01 & 0.21 & & 15.57 &  0.69 &  7.10 &  0.31 &  323 \\
288-G25$^1$ &  3.46 & 0.41 &  19.18 &  0.37 &  2.28 & 0.04 & & 17.70 &  0.34 &  2.10 & 0.04 & &   --- &   --- &   --- &   --- &  --- \\
311-G12 &  6.05 & 0.28 &  28.53 &  0.38 &  1.18 & 0.02 & & 28.74 &  0.46 &  1.18 & 0.02 & & 29.26 &  0.85 &  1.21 &  0.04 &  --- \\
315-G20 &  4.01 & 0.88 &  22.67 &  2.48 &  4.99 & 0.55 & & 19.36 &  1.30 &  4.26 & 0.29 & & 18.16 &  5.52 &  4.00 &  1.21 &  --- \\
321-G10 &  2.73 & 0.38 &  14.81 &  0.38 &  2.03 & 0.05 & & 13.80 &  0.27 &  1.89 & 0.04 & &   --- &   --- &   --- &   --- &  145 \\
322-G73 &  5.01 & 1.22 &  16.65 &  0.98 &  2.87 & 0.17 & & 11.59 &  0.70 &  1.99 & 0.12 & &   --- &   --- &   --- &   --- &  --- \\
322-G87 &  2.68 & 0.35 &  17.01 &  0.52 &  2.21 & 0.07 & & 14.69 &  0.37 &  1.91 & 0.05 & &   --- &   --- &   --- &   --- &  149 \\
340-G08 &  2.23 & 0.32 &  18.06 &  0.48 &  2.63 & 0.07 & & 13.40 &  0.23 &  1.95 & 0.03 & &   --- &   --- &   --- &   --- & ~~99 \\
340-G09 &  4.24 & 0.50 &  19.08 &  0.80 &  2.26 & 0.09 & & 16.87 &  0.61 &  2.00 & 0.07 & & 12.28 &  0.77 &  1.45 &  0.09 & ~~96 \\
358-G26 &  6.19 & 0.46 &  12.84 &  0.20 &  0.95 & 0.01 & & 13.10 &  0.20 &  0.97 & 0.01 & &   --- &   --- &   --- &   --- &  --- \\
358-G29$^1$ &  6.95 & 0.56 &  12.64 &  0.30 &  1.02 & 0.02 & & 12.88 &  0.32 &  1.04 & 0.03 & & 14.73 &  0.90 &  1.19 &  0.07 &  160 \\
377-G07 &  3.59 & ---  &  41.44 &  7.12 &   --- &  --- & & 33.85 &  4.65 &   --- &  --- & &   --- &   --- &   --- &   --- &  --- \\
383-G05 &  8.85 & 1.46 &  36.78 &  3.17 &  6.06 & 0.52 & & 28.18 &  1.39 &  4.65 & 0.23 & & 21.36 &  1.09 &  3.52 &  0.18 &  --- \\
416-G25$^1$ &  3.47 & 0.96 &  12.94 &  0.66 &  3.59 & 0.18 & & 11.28 &  0.49 &  3.13 & 0.13 & &  8.82 &  0.21 &  2.45 &  0.06 &  209 \\
435-G14 &  2.90 & 0.52 &  25.71 &  1.56 &  4.62 & 0.28 & & 16.66 &  0.55 &  3.00 & 0.10 & & 15.88 &  0.82 &  2.86 &  0.15 &  162 \\
435-G25 &  4.98 & 0.57 &  88.08 &  4.63 & 10.16 & 0.53 & & 51.25 &  2.17 &  5.91 & 0.25 & & 44.09 &  1.76 &  5.09 &  0.20 &  201 \\
435-G50 &  2.32 & 0.27 &  18.22 &  0.74 &  2.14 & 0.09 & & 14.40 &  0.46 &  1.69 & 0.05 & &   --- &   --- &   --- &   --- & ~~79 \\
437-G62 &  5.31 & 0.66 &  30.42 &  1.67 &  3.77 & 0.21 & & 25.77 &  0.84 &  3.20 & 0.10 & & 21.52 &  0.64 &  2.67 &  0.08 &  --- \\
444-G21 &  2.33 & 0.43 &  14.60 &  2.47 &  2.72 & 0.46 & & 15.94 &  1.82 &  2.97 & 0.34 & &   --- &   --- &   --- &   --- &  109 \\
446-G18 &  2.27 & 0.55 &  22.38 &  0.64 &  5.38 & 0.15 & & 16.60 &  0.36 &  3.99 & 0.09 & & 14.27 &  0.43 &  3.43 &  0.10 &  190 \\
446-G44 &  2.56 & 0.40 &  52.00 &  5.05 &  8.17 & 0.79 & & 27.88 &  1.08 &  4.38 & 0.17 & & 18.91 &  0.45 &  2.97 &  0.07 &  151 \\
460-G31 &  3.74 & 0.97 &  38.36 &  4.72 &  9.95 & 1.22 & & 26.26 &  1.92 &  6.81 & 0.50 & & 19.01 &  1.35 &  4.93 &  0.35 &  225 \\
487-G02$^2$ &  4.71 & 0.48 &  24.09 &  0.42 &  2.44 & 0.04 & & 21.66 &  0.39 &  2.19 & 0.04 & & 22.92 &  0.57 &  2.32 &  0.06 &  154 \\
500-G24 &  5.09 & 0.53 &  17.00 &  0.21 &  1.76 & 0.02 & & 17.35 &  0.31 &  1.80 & 0.03 & & 19.73 &  0.79 &  2.05 &  0.08 &  --- \\
505-G03 &  3.14 & 0.20 &  55.78 &  8.44 &  3.52 & 0.53 & & 37.46 &  3.79 &  2.36 & 0.24 & &   --- &   --- &   --- &   --- & ~~89 \\
506-G02 &  3.12 & 0.75 &  31.64 &  5.15 &  7.59 & 1.24 & & 22.44 &  2.98 &  5.39 & 0.71 & &   --- &   --- &   --- &   --- &  208 \\
509-G19 &  3.64 & 1.86 &  23.03 &  1.88 & 11.74 & 0.96 & & 17.39 &  1.14 &  8.86 & 0.58 & & 18.52 &  2.07 &  9.44 &  1.06 &  --- \\
531-G22 &  3.13 & 0.55 &  25.81 &  0.84 &  4.55 & 0.15 & & 21.27 &  0.82 &  3.75 & 0.14 & &   --- &   --- &   --- &   --- &  177 \\
555-G36$^2$ &  3.31 & ---  &    --- &   --- &   --- &  --- & &   --- &   --- &   --- &  --- & &   --- &   --- &   --- &   --- &  --- \\
564-G27 &  3.26 & 0.46 &  67.07 &  5.21 &  9.43 & 0.73 & & 37.96 &  1.08 &  5.34 & 0.15 & & 40.59 &  1.85 &  5.71 &  0.26 &  162 \\
575-G61 &  3.27 & 0.31 &  25.25 &  1.38 &  2.39 & 0.13 & & 18.40 &  0.36 &  1.74 & 0.03 & &   --- &   --- &   --- &   --- & ~~65 \\
\hline
\end{tabular}
\end{center}
\begin{flushleft}
{\sc Notes:}\\
$^1$ These galaxies show an additional component, a ``lens'', with a
luminosity profile in between the bulge and the disk component.  The
fits were applied to the following fitting ranges: ESO033-G22 --- 2--4
$h_{R,I}$; ESO288-G25 -- 2--6 $h_{R,I}$; ESO358-G29 and ESO416-G25 ---
2--6 $h_{R,K}$.  \\
$^2$ These galaxies are either greatly affected by foreground stars
(ESO138-G14, fits between 2 and 4 $h_{R,K}$, and ESO555-G36), have a
disk that cannot be approximated well by an exponential luminosity
decline (ESO487-G02, fits between 1 and 3 $h_{R,K}$), or show a very
irregular radial profile (ESO074-G15, fits between 0 and 2 $h_{R,I}$)
\end{flushleft}
\end{table*}
}

\begin{figure*}
\hspace*{0.7cm}
\psfig{figure=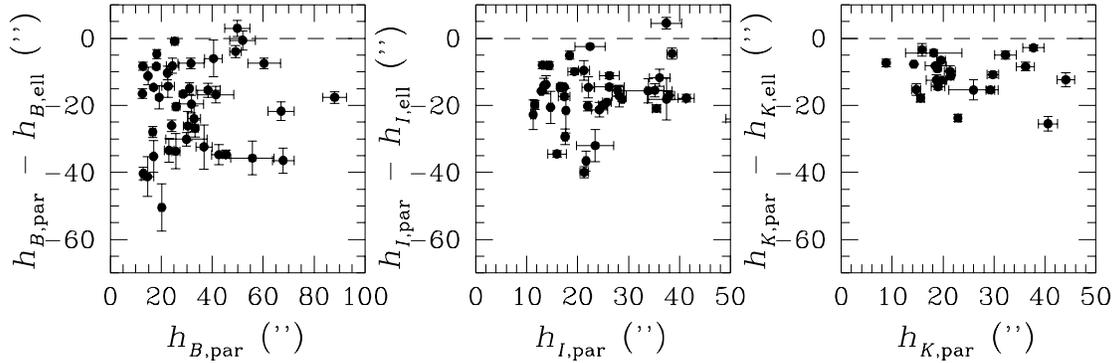,width=16cm}
\vspace*{-10.5cm}
\caption[]{\label{comparison.fig}Comparison between the scale lengths
determined using elliptically averaged luminosity profiles ($h_{\rm
<band>,ell}$) and profiles extracted parallel to the galaxies' major
axes ($h_{\rm <band>,par}$).  The dashed lines indicate the locus of the
data points if there were no difference between the two methods.}
\end{figure*}

In Table \ref{global1.tab} we present the {\it B, I,} and {\it K}-band
scale lengths of our sample galaxies determined from luminosity profiles
extracted at positions parallel to the major axes.  To obtain these
scale lengths, we applied a least-squares minimization algorithm to the
luminosity profiles between 1 and 4 {\it K}-band radial scale lengths,
unless indicated otherwise.  The errors we quote are realistic rather
than statistical in the sense that they represent the possible
deviations from the mean if the radial fitting range is varied. 

To obtain luminosity profiles representative of the light distributions
parallel to the galaxies' major axes, we decided to extract them at
those {\it z} distances, on the least dusty side of the planes, where
the effects of dust extinction are greatly reduced.  In choosing these
distances we were limited by the signal-to-noise ratio in the {\it K}
band.  The best choice for the {\it z} distance to extract the profiles
from turned out to be between 1.0 and 1.5 vertical scale heights. 
Although the influence of the central dust lane is greatly reduced at
these distances from the planes, it is not (yet) completely negligible.

{
\begin{table}

\caption[ ]{\label{global2.tab}{\bf Extrapolated central surface
brightnesses}\\ All values were obtained using a radial fitting range
between 1 and 4 {\it K}-band scale lengths, $h_{R,K}$ ({\it I}-band if
no {\it K}-band data was available).}

\begin{center}
\tabcolsep=1.5mm

\begin{tabular}{crrrrrr}
\hline
\multicolumn{1}{c}{Galaxy} & \multicolumn{1}{c}{$\mu_{0,B}$} &
\multicolumn{1}{c}{$\pm$} & \multicolumn{1}{c}{$\mu_{0,I}$} &
\multicolumn{1}{c}{$\pm$} & \multicolumn{1}{c}{$\mu_{0,K}$} &
\multicolumn{1}{c}{$\pm$} \\
\cline{2-7}
\noalign{\vspace{2pt}}
\multicolumn{1}{c}{(ESO-LV)} & \multicolumn{6}{c}{({\it B, I, K} mag
arcsec$^{-2}$)} \\ 
\multicolumn{1}{c}{(1)} & \multicolumn{1}{c}{(2)} &
\multicolumn{1}{c}{(3)} & \multicolumn{1}{c}{(4)} &
\multicolumn{1}{c}{(5)} & \multicolumn{1}{c}{(6)} &
\multicolumn{1}{c}{(7)} \\
\hline 
026-G06 & 22.16 & 0.15 & 20.11 & 0.14 & 17.54 & 0.10 \\
033-G22 & 22.37 & 0.41 & 20.66 & 0.51 &   --- &  --- \\
041-G09 & 22.28 & 0.22 & 19.44 & 0.10 & 16.45 & 0.05 \\
074-G15 & 21.13 & 0.18 & 19.55 & 0.08 &   --- &  --- \\
138-G14 & 20.70 & 0.10 & 19.51 & 0.07 & 18.34 & 0.52 \\
141-G27 & 21.55 & 0.06 & 19.66 & 0.06 & 17.11 & 0.03 \\
142-G24 & 20.42 & 0.09 & 19.12 & 0.17 & 17.37 & 0.12 \\
157-G18 & 21.25 & 0.08 & 19.29 & 0.07 & 16.02 & 0.04 \\
201-G22 & 19.36 & 0.67 & 18.28 & 0.11 &   --- &  --- \\
202-G35 & 22.81 & 0.28 & 22.02 & 0.87 &   --- &  --- \\
235-G53 & 21.64 & 0.13 & 19.16 & 0.42 &   --- &  --- \\
240-G11 & 21.94 & 0.17 & 18.82 & 0.15 & 15.77 & 0.13 \\
263-G15 & 22.08 & 0.23 & 19.49 & 0.24 &   --- &  --- \\
263-G18 & 21.42 & 0.17 & 18.83 & 0.08 &   --- &  --- \\
269-G15 & 21.54 & 0.17 & 19.33 & 0.11 & 16.34 & 0.14 \\
286-G18 & 20.03 & 0.18 & 18.53 & 0.30 &   --- &  --- \\
288-G25 & 20.14 & 0.10 & 18.04 & 0.04 & 15.53 & 0.19 \\
311-G12 & 23.10 & 0.21 & 19.72 & 0.32 & 15.98 & 0.12 \\
315-G20 & 19.92 & 0.18 & 18.51 & 0.36 &   --- &  --- \\
321-G10 & 21.43 & 0.32 & 19.75 & 0.28 &   --- &  --- \\
322-G73 & 21.44 & 0.45 & 18.95 & 0.45 &   --- &  --- \\
322-G87 & 21.12 & 0.10 & 20.18 & 0.17 &   --- &  --- \\
340-G08 & 20.34 & 0.43 & 21.45 & 0.17 & 19.19 & 0.59 \\
340-G09 & 18.21 & 0.06 & 16.16 & 0.05 &   --- &  --- \\
358-G26 & 19.13 & 0.19 & 17.36 & 0.22 & 15.66 & 0.36 \\
358-G29 & 22.22 & 0.62 & 20.95 & 0.52 &   --- &  --- \\
377-G07 & 22.60 & 0.13 & 19.92 & 0.12 & 16.24 & 0.10 \\
383-G05 & 21.59 & 0.38 & 19.47 & 0.30 & 16.18 & 0.65 \\
416-G25 & 20.65 & 0.09 & 18.44 & 0.08 & 15.84 & 0.26 \\
435-G14 & 21.69 & 0.22 & 18.73 & 0.24 & 15.49 & 0.11 \\
435-G25 & 20.51 & 0.05 & 19.73 & 0.05 &   --- &  --- \\
435-G50 & 20.37 & 0.33 & 18.13 & 0.05 & 15.20 & 0.09 \\
437-G62 & 20.86 & 0.39 & 20.57 & 0.26 &   --- &  --- \\
444-G21 & 20.90 & 0.09 & 18.78 & 0.13 & 15.40 & 0.09 \\
446-G18 & 20.96 & 0.09 & 18.74 & 0.05 & 15.79 & 0.03 \\
446-G44 & 22.49 & 0.20 & 19.82 & 0.09 & 16.93 & 0.12 \\
460-G31 & 20.53 & 0.06 & 18.64 & 0.07 & 15.83 & 0.12 \\
487-G02 & 19.54 & 0.04 & 17.84 & 0.06 & 15.79 & 0.06 \\
500-G24 & 20.58 & 0.08 & 19.39 & 0.18 &   --- &  --- \\
505-G03 & 20.46 & 0.11 & 18.39 & 0.13 &   --- &  --- \\
506-G02 & 21.59 & 0.09 & 18.83 & 0.07 & 15.61 & 0.04 \\
509-G19 & 20.77 & 0.23 & 19.54 & 0.24 &   --- &  --- \\
531-G22 & 21.73 & 0.11 & 20.54 & 0.15 &   --- &  --- \\
555-G36 & 23.26 & 0.37 & 19.60 & 0.30 & 16.96 & 0.14 \\
564-G27 & 21.34 & 0.11 & 20.29 & 0.13 &   --- &  --- \\
\hline
\end{tabular}
\end{center}
\end{table}
}

\begin{figure}
\psfig{figure=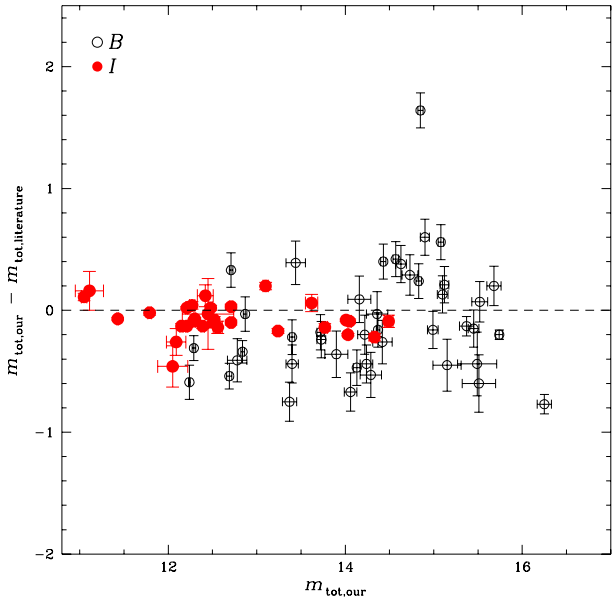,width=8.7cm}
\caption[]{\label{magnitudes.fig}Comparison of our {\it B}- and {\it
I}-band magnitudes to those published in the RC3 ({\it B} band) and by
Mathewson et al. (1992, {\it I} band). The dashed line indicates the
locus of equality.}
\end{figure}

We used heliocentric velocities obtained by Mathewson et al.  (1992) for
the majority of our sample galaxies to base the calculation of our
distance-dependent parameters on.  The effects of the large-scale Hubble
expansion were corrected for using Richter et al.'s (1987) formalism,
which supposedly yields highly reliable values (Schmidt \& Boller 1992). 

As already touched upon in Sect.  \ref{hlength.sect}, since the effects
that influence the determination of the scale lengths from elliptically
averaged luminosity profiles of edge-on disc galaxies are not well
understood, and may even be dependent on galaxy type, they should
preferably not be used in an analysis involving global scale parameters. 
To assess the importance of the effects discussed in Sect. 
\ref{hlength.sect}, in Fig.  \ref{comparison.fig} we compare the scale
lengths derived from our sample galaxies using both methods.  We notice
clear systematic effects in all passbands, in the sense that the scale
lengths obtained from the ``elliptical'' profiles are generally larger
than those determined from the ``parallel'' profiles.  The effect is
smallest in the near-infrared, indicating that it is probably caused by
dust contamination. 

Although the ellipse fits should preferably not be used to derive scale
parameters from the observed galaxy images, they are useful in providing
estimates for the (edge-on) central surface brightnesses of galaxy
discs.  In Table \ref{global2.tab} we tabulate the extrapolated edge-on
disc central surface brightnesses. 

The calibration of the observations was discussed in de Grijs et al. 
(1997).  It was shown that our data agree sufficiently well in detail
with those published previously.  Here we present, in Table
\ref{photprop.tab}, the apparent magnitudes in the {\it B, I,} and {\it
K} bands of the total galaxies as well as those of the disc components
only.  The apparent disc magnitudes are based on model discs obtained
from ellipse fits to the outer galaxy isophotes. 

In Fig.  \ref{magnitudes.fig} we compare the (total) apparent {\it B}
and {\it I}-band magnitudes with literature values; we took the apparent
{\it B}-band magnitudes from de Vaucouleurs et al.  (1991, RC3), and the
apparent {\it I}-band magnitudes from the observations by Mathewson et
al.  (1992).  Especially our determinations of the {\it I}-band apparent
magnitudes agree rather well with those determined by Mathewson et al. 
(1992): the mean deviation between our and Mathewson et al.'s (1992)
magnitude determinations is $-0.07 \pm 0.13$ mag (based on a comparison
of 30 galaxies); in the {\it B} band a larger scatter (of 0.45 mag, for
43 galaxies) between our and the RC3 determinations was found, although
we do not detect a systematic difference between our apparent {\it
B}-band magnitudes and the RC3 magnitudes (the mean deviation equals
$-0.09$ mag). 

In Table \ref{photprop.tab}, we also present the absolute magnitudes in
the {\it I} and {\it K} bands, $M_I$ and $M_K$.  We did not include the
{\it B}-band absolute magnitudes, as those are too heavily affected by
the internal dust extinction to be reliable. 

{
\begin{table*}

\caption[ ]{\label{photprop.tab}{\bf Photometric characteristics of the
sample galaxies}\\ Columns: (1) Galaxy name (ESO-LV); (2)--(5),
(6)--(9), (10)--(13) Apparent magnitudes of the total galaxy and the
disk component only, and the 1$\sigma$ observational errors, for the
{\it B, I} and {\it K} bands (both determined in Chapter 6),
respectively.  The systematic errors due to variations in the
photometric accuracy are of order 0.07 mag in {\it B}, 0.04 mag in {\it
I}, and 0.08 mag in {\it K}; (14)--(16) Galactic extinction in the {\it
B} band (taken from the RC3), the {\it I} and the {\it K} bands,
respectively; (17)--(18) {\it I} and {\it K}-band absolute magnitudes,
corrected for Galactic extinction (The {\it B}-band absolute magnitudes
are generally heavily affected by dust extinction and have therefore not
been included).}

\begin{center} 
\tabcolsep=1mm

\begin{tabular}{rrrrrcrrrrcrrrrccrrcc}
\hline
\noalign{\vspace*{2pt}}
\multicolumn{1}{c}{Galaxy} & \multicolumn{4}{c}{{\it B}-band magnitudes}
& & \multicolumn{4}{c}{{\it I}-band magnitudes} & &
\multicolumn{4}{c}{{\it K}-band magnitudes} & &
\multicolumn{1}{c}{$A_{G,B}$} & \multicolumn{1}{c}{$A_{G,I}$} &
\multicolumn{1}{c}{$A_{G,K}$} & \multicolumn{1}{c}{$M_I^0$} &
\multicolumn{1}{c}{$M_K^0$} \\ 
\cline{2-5} \cline{7-10} \cline{12-15} \cline{17-21}
\multicolumn{1}{c}{(ESO)} & \multicolumn{1}{c}{total} &
\multicolumn{1}{c}{$\pm$} & \multicolumn{1}{c}{disk} &
\multicolumn{1}{c}{$\pm$} & & \multicolumn{1}{c}{total} &
\multicolumn{1}{c}{$\pm$} & \multicolumn{1}{c}{disk} &
\multicolumn{1}{c}{$\pm$} & & \multicolumn{1}{c}{total} &
\multicolumn{1}{c}{$\pm$} & \multicolumn{1}{c}{disk} &
\multicolumn{1}{c}{$\pm$} & & \multicolumn{5}{c}{({\it B, I, K} mag)} \\
\multicolumn{1}{c}{(1)}  & \multicolumn{1}{c}{(2)} &
\multicolumn{1}{c}{(3)}  & \multicolumn{1}{c}{(4)} &
\multicolumn{1}{c}{(5)}  & & \multicolumn{1}{c}{(6)} &
\multicolumn{1}{c}{(7)}  & \multicolumn{1}{c}{(8)} &
\multicolumn{1}{c}{(9)}  & & \multicolumn{1}{c}{(10)} &
\multicolumn{1}{c}{(11)} & \multicolumn{1}{c}{(12)} &
\multicolumn{1}{c}{(13)} & & \multicolumn{1}{c}{(14)} &
\multicolumn{1}{c}{(15)} & \multicolumn{1}{c}{(16)} &
\multicolumn{1}{c}{(17)} & \multicolumn{1}{c}{(18)} \\
\hline 
026-G06 & 15.68 & 0.08 & 15.80 & 0.08 & & 13.24 & 0.02 & 13.48 & 0.08 & & 11.50 & 0.09 & 11.69 & 0.10 & & 0.47 & 0.17 & 0.04 & --18.76 & --20.37 \\
033-G22 & 15.37 & 0.08 & 15.51 & 0.03 & & 14.03 & 0.01 & 14.33 & 0.06 & & ---   & ---  & ---   & ---  & & 0.43 & 0.16 & 0.04 & --19.10 &  ---    \\
041-G09 & 13.52 & 0.01 & 13.52 & 0.01 & & 12.21 & 0.01 & 12.21 & 0.03 & & ---   & ---  & ---   & ---  & & 0.62 & 0.23 & 0.05 & --20.78 &  ---    \\
138-G14 & 14.85 & 0.03 & 15.20 & 0.05 & & 14.01 & 0.01 & 14.37 & 0.03 & & ---   & ---  & ---   & ---  & & 0.60 & 0.22 & 0.05 & --18.82 &  ---    \\
141-G27 & 14.57 & 0.03 & 14.57 & 0.03 & & 12.71 & 0.02 & 12.83 & 0.04 & & 10.57 & 0.12 & 10.57 & 0.12 & & 0.21 & 0.08 & 0.02 & --18.18 & --20.26 \\
142-G24 & 14.43 & 0.03 & 14.43 & 0.03 & & 12.30 & 0.01 & 12.30 & 0.03 & & 10.56 & 0.16 & 10.56 & 0.16 & & 0.25 & 0.09 & 0.02 & --19.33 & --21.00 \\
157-G18 & 13.72 & 0.04 & 13.72 & 0.04 & & 12.29 & 0.02 & 12.29 & 0.02 & &  9.79 & 0.50 &  9.79 & 0.50 & & 0.00 & 0.00 & 0.00 & --18.35 & --20.85 \\
201-G22 & 14.06 & 0.07 & 14.06 & 0.07 & & 13.10 & 0.03 & 13.10 & 0.05 & & 10.34 & 0.16 & 10.34 & 0.16 & & 0.00 & 0.00 & 0.00 & --20.13 & --22.89 \\
202-G35 & 12.71 & 0.02 & 12.71 & 0.02 & & 11.79 & 0.02 & 11.79 & 0.04 & & ---   & ---  & ---   & ---  & & 0.00 & 0.00 & 0.00 & --19.13 &  ---    \\
235-G53 & 14.37 & 0.02 & 15.46 & 0.06 & & 12.45 & 0.01 & 13.19 & 0.04 & & ---   & ---  & ---   & ---  & & 0.09 & 0.03 & 0.01 & --21.10 &  ---    \\
240-G11 & 13.44 & 0.11 & 13.66 & 0.11 & & 11.11 & 0.16 & 11.50 & 0.06 & & ---   & ---  & ---   & ---  & & 0.04 & 0.01 & 0.00 & --21.55 &  ---    \\
263-G15 & 14.22 & 0.08 & 14.22 & 0.08 & & 11.19 & 0.01 & 11.19 & 0.01 & &  9.09 & 0.15 &  9.09 & 0.15 & & 1.24 & 0.45 & 0.11 & --20.98 & --22.74 \\
263-G18 & 14.29 & 0.12 & 14.33 & 0.12 & & 11.70 & 0.02 & 13.22 & 0.11 & & ---   & ---  & ---   & ---  & & 0.62 & 0.23 & 0.05 & --21.49 &  ---    \\
269-G15 & 14.16 & 0.13 & 14.16 & 0.13 & & 12.15 & 0.02 & 12.15 & 0.04 & & ---   & ---  & ---   & ---  & & 0.47 & 0.17 & 0.04 & --20.65 &  ---    \\
286-G18 & 14.90 & 0.05 & 14.90 & 0.05 & & 12.21 & 0.01 & 12.21 & 0.03 & & 10.45 & 0.38 & 10.45 & 0.38 & & 0.07 & 0.03 & 0.01 & --22.69 & --24.43 \\
288-G25 & 13.73 & 0.05 & 13.73 & 0.05 & & 11.62 & 0.01 & 11.62 & 0.03 & & ---   & ---  & ---   & ---  & & 0.00 & 0.00 & 0.00 & --20.33 &  ---    \\
311-G12 & 12.24 & 0.01 & 13.32 & 0.03 & & 10.04 & 0.01 & 10.85 & 0.03 & &  7.35 & 0.20 &  8.36 & 0.61 & & 1.49 & 0.54 & 0.13 & --20.15 & --22.43 \\
315-G20 & 16.25 & 0.08 & 17.56 & 0.40 & & 12.62 & 0.01 & 13.76 & 0.07 & &  9.19 & 0.78 & 11.03 & 0.98 & & 0.00 & 0.00 & 0.00 & --20.67 & --24.10 \\
321-G10 & 13.90 & 0.13 & 14.84 & 0.07 & & 12.05 & 0.17 & 12.05 & 0.15 & & ---   & ---  & ---   & ---  & & 0.35 & 0.13 & 0.03 & --20.34 &  ---    \\
322-G73 & 13.37 & 0.08 & 13.68 & 0.09 & & 11.84 & 0.02 & 12.26 & 0.02 & & ---   & ---  & ---   & ---  & & 0.52 & 0.19 & 0.04 & --21.10 &  ---    \\
322-G87 & 14.24 & 0.07 & 14.24 & 0.07 & & 12.09 & 0.11 & 12.09 & 0.11 & & ---   & ---  & ---   & ---  & & 0.49 & 0.18 & 0.04 & --20.23 &  ---    \\
340-G08 & 15.52 & 0.09 & 16.22 & 0.04 & & 13.77 & 0.03 & 13.95 & 0.04 & & ---   & ---  & ---   & ---  & & 0.16 & 0.06 & 0.01 & --18.68 &  ---    \\
340-G09 & 15.08 & 0.03 & 15.08 & 0.03 & & 13.62 & 0.07 & 13.62 & 0.07 & & 11.34 & 0.63 & 11.34 & 0.63 & & 0.20 & 0.07 & 0.02 & --18.39 & --20.62 \\
358-G26 & 12.87 & 0.02 & 13.11 & 0.02 & & 11.68 & 0.01 & 12.84 & 0.04 & & ---   & ---  & ---   & ---  & & 0.00 & 0.00 & 0.00 & --19.23 &  ---    \\
358-G29 & 12.29 & 0.02 & 13.14 & 0.04 & & 10.37 & 0.01 & 11.08 & 0.04 & &  8.26 & 0.29 &  9.16 & 0.74 & & 0.00 & 0.00 & 0.00 & --20.73 & --22.84 \\
377-G07 & 15.49 & 0.22 & 15.49 & 0.22 & & 13.49 & 0.07 & 13.49 & 0.09 & & ---   & ---  & ---   & ---  & & 0.34 & 0.12 & 0.03 &     --- &  ---    \\
383-G05 & 14.63 & 0.06 & 14.87 & 0.08 & & 11.94 & 0.02 & 12.62 & 0.05 & &  8.95 & 1.15 &  9.72 & 0.24 & & 0.16 & 0.06 & 0.01 & --20.78 & --23.72 \\
416-G25 & 14.42 & 0.11 & 14.65 & 0.04 & & 12.52 & 0.03 & 12.81 & 0.05 & & 10.35 & 0.76 & 11.38 & 1.12 & & 0.03 & 0.01 & 0.00 & --21.28 & --23.44 \\
435-G14 & 14.73 & 0.09 & 14.73 & 0.09 & & 12.56 & 0.05 & 12.56 & 0.08 & & 10.12 & 0.37 & 10.12 & 0.37 & & 0.18 & 0.07 & 0.02 & --20.36 & --22.75 \\
435-G25 & 12.78 & 0.11 & 13.23 & 0.14 & & 11.05 & 0.03 & 11.54 & 0.07 & &  8.42 & 0.41 &  8.63 & 0.52 & & 0.25 & 0.09 & 0.02 & --20.92 & --23.48 \\
435-G50 & 15.74 & 0.04 & 15.74 & 0.04 & & 14.33 & 0.03 & 14.33 & 0.05 & & ---   & ---  & ---   & ---  & & 0.29 & 0.11 & 0.02 & --17.70 &  ---    \\
437-G62 & 12.69 & 0.03 & 13.47 & 0.07 & & 10.72 & 0.01 & 11.28 & 0.03 & &  8.34 & 0.19 &  8.65 & 0.11 & & 0.31 & 0.11 & 0.03 & --21.43 & --23.73 \\
444-G21 & 15.15 & 0.16 & 15.15 & 0.16 & & 14.05 & 0.02 & 14.05 & 0.03 & &  ---  & ---  & ---   & ---  & & 0.25 & 0.09 & 0.02 & --18.96 &  ---    \\
446-G18 & 15.12 & 0.05 & 15.32 & 0.06 & & 12.71 & 0.02 & 13.21 & 0.04 & & 10.17 & 0.42 & 10.84 & 0.42 & & 0.22 & 0.08 & 0.02 & --20.85 & --23.33 \\
446-G44 & 14.83 & 0.03 & 14.83 & 0.03 & & 12.51 & 0.04 & 12.51 & 0.06 & & 10.23 & 0.66 & 10.23 & 0.66 & & 0.29 & 0.11 & 0.02 & --20.15 & --22.34 \\
460-G31 & 14.99 & 0.06 & 15.39 & 0.09 & & 12.48 & 0.02 & 13.28 & 0.03 & & 10.02 & 0.56 & 10.55 & 0.60 & & 0.71 & 0.26 & 0.06 & --21.42 & --23.68 \\
487-G02 & 13.40 & 0.03 & 13.66 & 0.04 & & 11.43 & 0.01 & 11.49 & 0.03 & &  8.84 & 0.39 &  9.05 & 0.82 & & 0.04 & 0.01 & 0.00 & --20.18 & --22.76 \\
500-G24 & 12.84 & 0.02 & 13.25 & 0.03 & & 10.92 & 0.01 & 11.31 & 0.03 & & ---   & ---  & ---   & ---  & & 0.34 & 0.12 & 0.03 & --20.85 &  ---    \\
505-G03 & 13.40 & 0.07 & 13.40 & 0.07 & & 12.39 & 0.02 & 12.39 & 0.04 & & ---   & ---  & ---   & ---  & & 0.25 & 0.09 & 0.02 & --18.27 &  ---    \\
506-G02 & 14.13 & 0.04 & 14.37 & 0.27 & & 12.45 & 0.29 & 13.06 & 0.11 & & ---   & ---  & ---   & ---  & & 0.46 & 0.17 & 0.04 & --21.19 &  ---    \\
509-G19 & 15.10 & 0.06 & 15.10 & 0.06 & & 12.32 & 0.01 & 12.32 & 0.04 & &  9.58 & 0.44 &  9.58 & 0.44 & & 0.21 & 0.08 & 0.02 & --22.87 & --25.55 \\
531-G22 & 13.96 & 0.08 & 13.96 & 0.08 & & 12.27 & 0.03 & 12.27 & 0.06 & & ---   & ---  & ---   & ---  & & 0.10 & 0.04 & 0.01 & --20.58 &  ---    \\
555-G36 & 15.45 & 0.06 & 15.45 & 0.06 & & 14.18 & 0.03 & 14.18 & 0.06 & & ---   & ---  & ---   & ---  & & 0.47 & 0.17 & 0.04 &     --- &  ---    \\
564-G27 & 14.36 & 0.12 & 14.56 & 0.12 & & 12.42 & 0.09 & 13.02 & 0.11 & &  9.72 & 0.34 &  9.72 & 0.34 & & 0.56 & 0.20 & 0.05 & --20.09 & --22.64 \\
575-G61 & 15.51 & 0.19 & 15.51 & 0.19 & & 14.49 & 0.05 & 14.49 & 0.08 & & ---   & ---  & ---   & ---  & & 0.25 & 0.09 & 0.02 & --17.05 &  ---    \\
\hline
\end{tabular}
\end{center}
\end{table*}
}

\subsection{Scale parameter ratios}
\label{scalepars.sect}

\begin{figure}
\psfig{figure=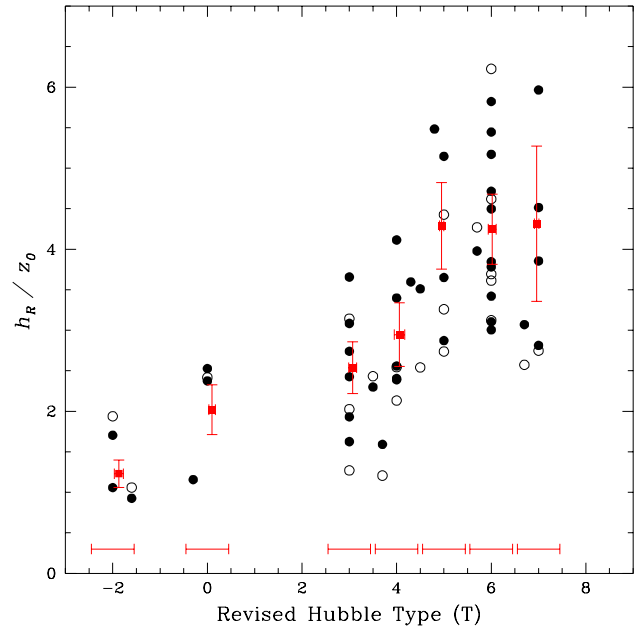,width=9cm}
\caption[]{\label{ratio_type.fig}Dependence of the $h_R/z_0$ ratio on
galaxy type for both {\it I}-band data (filled dots) and {\it K}-band
observations (open circles).  The filled squares show the {\it I}-band
ratios averaged over the type bins indicated by the horizontal bars; the
errors indicate the standard deviations of the distribution.}
\end{figure}

Fig.  \ref{ratio_type.fig} shows the dependence of the scale parameter
ratio on galaxy type; we have also plotted the average scale parameter
ratios per galaxy type.  We notice a correlation between the $h_R/z_0$
ratio and revised Hubble type, in the sense that galaxies become
systematically thinner when going from S0's to Sc's ($T=5$), whereas the
later types (Sd's, $T=7$) seem to be at least as thin as the Sc's. 

\begin{figure}
\psfig{figure=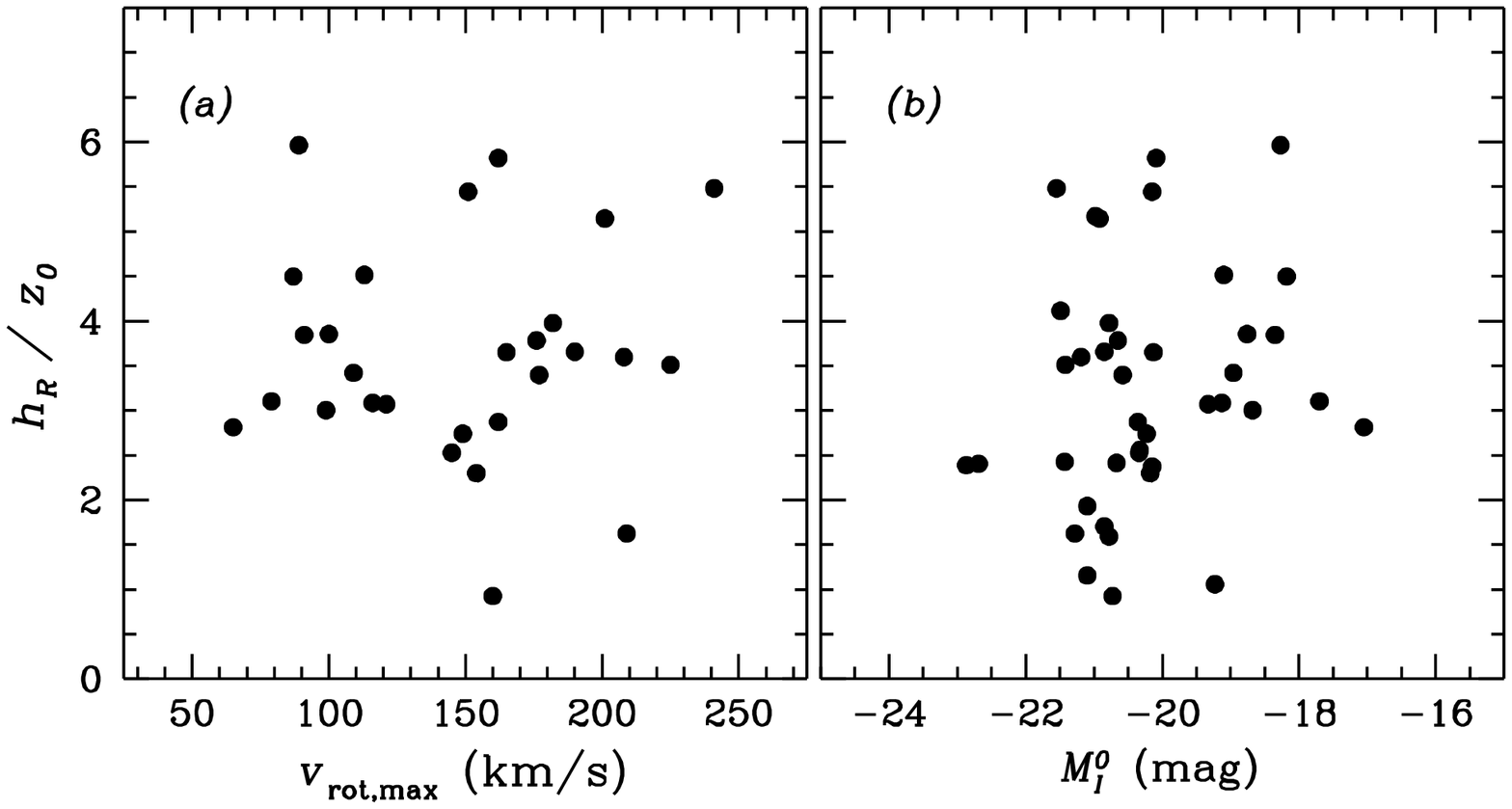,width=9cm}
\vspace{-4cm}
\caption[]{\label{scale_ratios.fig}{\it I}-band scale parameters versus
{\it (a)} the maximum rotation velocity, and {\it (b)} the absolute {\it
I}-band magnitudes.}
\end{figure}

Although a correlation between the $h_R/z_0$ ratio and the rotation
velocity was expected, based on the qualitative arguments presented in
Sect.  \ref{ratio.sec}, we do not find any evidence for such a
dependence (Fig.  \ref{scale_ratios.fig}a), nor do we find any
evidence for a relationship between the $h_R/z_0$ ratio and the absolute
magnitudes (Fig.  \ref{scale_ratios.fig}b).  In other words, the
theoretical prediction that the scale parameter should decrease rapidly
from faint galaxies to a constant level for normal and bright galaxies
(Bottema 1993) could not be confirmed observationally.  This implies
that either the relationship is weak, i.e., smaller than the
observational scatter, or the underlying assumptions (of, e.g., a
constant $(M/L)_B$, and a linear relationship between the old-disc
absolute magnitude and the disc vertical velocity dispersion) either
have a large intrinsic scatter themselves or are not valid. 

\subsection{Radial colour gradients}
\label{radgrads.sec}

\begin{figure*}
\hspace*{0.7cm}
\psfig{figure=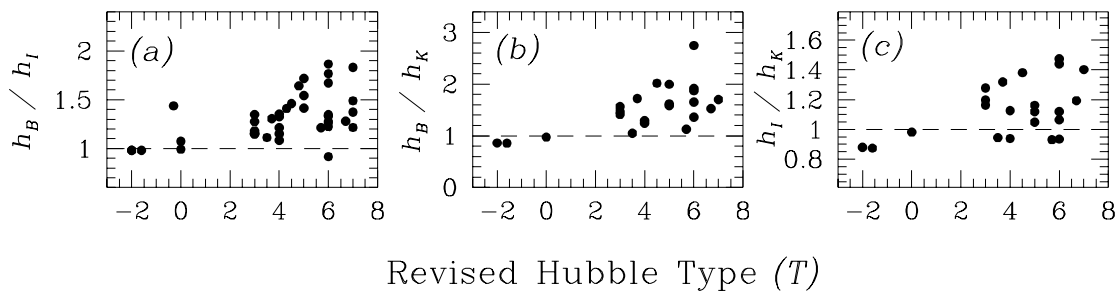,width=16cm}
\vspace*{-11.5cm}
\caption[]{\label{hratios.fig}Scale length ratios as a function of
revised Hubble type derived from the profiles extracted parallel to the
galaxies' major axes.  The dashed lines indicate the locus of galaxies
which do not show any scale length difference between passbands.}
\end{figure*}

As we argued in Sect.  \ref{colgrads.sect}, one can study radial colour
gradients within galaxies by comparing scale lengths determined in
different passbands.  In Fig.  \ref{hratios.fig} we show the scale
length ratios (indicating radial colour gradients) obtained for our
sample galaxies. 

In Table \ref{hratios.tab} we present the mean scale length ratios, both
for our total sample and for the galaxies of type $T > 2$.

{
\begin{table}

\caption[ ]{\label{hratios.tab}{\bf Scale length ratios of the sample
galaxies}\\
Columns: (1) Scale length ratio; (2) (Sub)sample used; (4) Number of
galaxies in the (sub)sample; (5) Resulting scale length ratio and
standard deviation.}

\begin{center}
\tabcolsep=1.5mm

\begin{tabular}{clcc}
\hline
Gradient  & \multicolumn{1}{c}{Galaxy} & Number of & \multicolumn{1}{c}{Mean} \\
          & \multicolumn{1}{c}{types}  & Galaxies & \multicolumn{1}{c}{Ratio} \\
\multicolumn{1}{c}{(1)} & \multicolumn{1}{c}{(2)} &
\multicolumn{1}{c}{(3)} & \multicolumn{1}{c}{(4)} \\
\hline
$h_B/h_I$ & $T > 2$   & 40 & 1.36 $\pm$ 0.22 \\
          & All types & 45 & 1.32 $\pm$ 0.24 \\
$h_B/h_K$ & $T > 2$   & 22 & 1.65 $\pm$ 0.41 \\
          & All types & 25 & 1.56 $\pm$ 0.46 \\
$h_I/h_K$ & $T > 2$   & 22 & 1.19 $\pm$ 0.17 \\
          & All types & 25 & 1.15 $\pm$ 0.19 \\
\hline
\end{tabular}
\end{center}
\end{table}
}

The main conclusion that we draw in de Grijs et al.  (1997), based on
the ``elliptical'' {\it I} versus {\it K}-band scale length ratios,
still holds firmly, namely that on average the colour gradients
indicated by the scale length ratios increase from type Sb ($T=3$) to at
least type Scd ($T=6$), as is best seen in Figs.  \ref{hratios.fig}a and
b.  For galaxy types later than Scd, the colour gradients seem to become
smaller again, although -- due to small-number statistics -- we cannot
draw any firm conclusions in that range of galaxy types. 

For the earliest-type sample galaxies ($T<2$) we find very small colour
gradients at these {\it z} distances.  It is, in fact, expected that the
intrinsic colour gradients at these heights above the planes are smaller
than those in the planes, where the young population contributes to the
luminosity.  At these {\it z} heights we are looking at the old-disc
population, which is very uniform in colour, as was argued in Sect.
\ref{edgeongrad.sect}. 

The results presented in this section are consistent with those
published previously (e.g., Elmegreen \& Elmegreen 1984; Peletier et
al. 1994, 1995a; de Jong 1996a; and others), as will be shown in Sect. 
\ref{radgrads.sect}. 

\subsection{Correlations between gross galaxy properties}

\begin{figure}
\psfig{figure=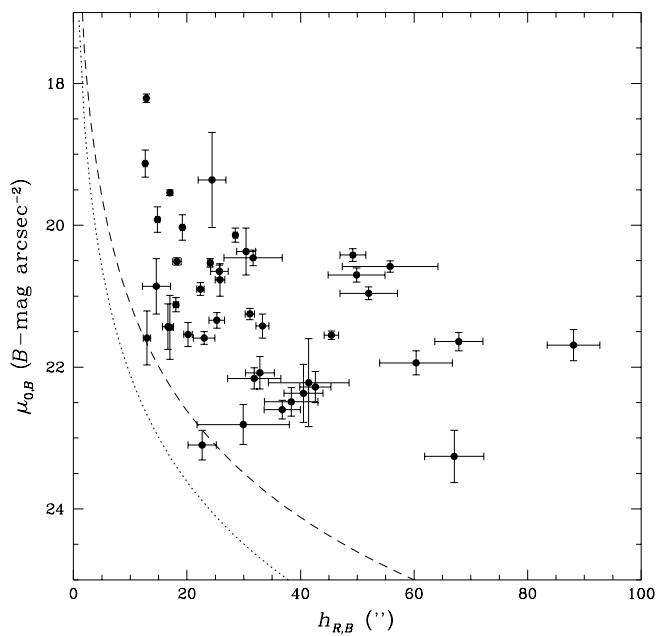,width=9cm}
\caption[]{\label{selection.fig}Illustration of the effects of sample
selection based on a diameter limit.  The dashed line represents a
diameter-selection limit (of $2.'2$) at 25 {\it B}-mag arcsec$^{-2}$;
the dotted line is the limit for a diameter-selection limit at 26 {\it
B}-mag arcsec$^{-2}$.}
\end{figure}

Since the sample we are dealing with is a diameter-selected sample of
edge-on disc galaxies, a correction has to be applied to the observed
parameters to represent a volume-limited sample (e.g., Davies 1990). 
Because the sample galaxies were selected to have a minimum blue major
axis diameter at 25 {\it B}-mag arcsec$^{-2}$, $D_{25}^B$, of $2.'2$, we
have created a selection bias against low surface brightness galaxies
and galaxies with small scale lengths.  The main implications of such a
selection bias have been discussed by de Jong (1996b).  In Fig. 
\ref{selection.fig} we show the implications of our selection bias with
respect to the disc properties of the observed sample galaxies in the
($h_R,\mu_0$) plane. 

The dashed and dotted lines in this figure represent the selection
limits for a limiting diameter of $2.'2$, using the 25 and 26 {\it
B}-mag arcsec$^{-2}$ isophotes, for comparison.  The diameter-selection
criterion used implies that our sample should only contain galaxies to
the right of the selection limit in Fig.  \ref{selection.fig}. 

Although our sample selection was based on the $D_{\rm orig}$ diameter
tabulated in the ESO-LV catalogue, this diameter was not determined at
exactly the 25 {\it B}-mag arcsec$^{-2}$ isophote (ESO-LV), but varies
with galaxy type.  For spiral galaxies the diameters correspond to
approximately 26 {\it B}-mag arcsec$^{-2}$ isophotes; for earlier types
the corresponding isophotes are in the range from 25.0--25.6 {\it B}-mag
arcsec$^{-2}$. 

As de Jong (1996b) remarks, since the number of galaxies in the sample
will decrease as $h_R^3$ if all galaxies have the same scale length, it
is not surprising that our sample does not contain galaxies with central
{\it B}-band surface brightnesses fainter than about 23 {\it B}-mag
arcsec$^{-2}$.

\subsubsection{Type dependences}

\begin{figure*}
\hspace*{0.7cm}
\psfig{figure=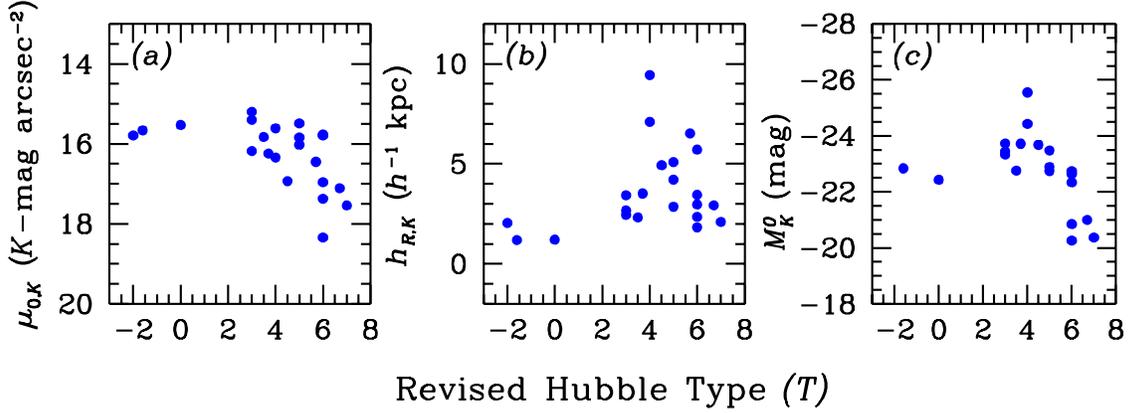,width=16cm}
\vspace*{-10cm}
\caption[]{\label{type.fig}Distribution of global {\it K}-band galaxy
properties as a function of galaxy type: {\it (a)} Central disc surface
brightnesses; {\it (b)} disc radial scale lengths (in $h^{-1}$ kpc);
{\it (c)} Absolute magnitudes of the sample galaxies.}
\end{figure*}

In Fig.  \ref{type.fig} we show the distributions of global {\it K}-band
galaxy parameters as a function of Hubble type.  We did not apply an
inclination correction to the galaxy parameters, but since we are
dealing with edge-on galaxies, inclination {\it differences} do not play
a significant role. 

The distribution of (edge-on) {\it K}-band disc central surface
brightnesses is rather flat (Fig.  \ref{type.fig}a), although with
a large scatter.  However, the latest-type sample galaxies ($T \ge 6$)
show an indication that their disc central surface brightnesses may be
fainter than those of the earlier types.  Most likely, this effect is
not the result of dust extinction, since the latest-type galaxies are
not likely to contain more dust than the earlier and intermediate types,
as was argued in Sect.  \ref{radgrads.sec}.  These results are
consistent with those of de Jong (1996b). 

Fig.  \ref{type.fig}b shows the distribution of the radial scale lengths
of the galaxy discs, in kpc; we do not notice any clear correlation with
galaxy type, apart from the observation that the earliest-type sample
galaxies apparently have the smallest scale lengths, with only a small
range of possible scale lengths compared to the range in the scale
length distribution for galaxies of type $T>2$.  De Jong (1996b) noticed
a possible lack of late-type galaxies with small scale lengths for his
sample of 86 face-on disc galaxies, but remarked that this could be a
selection effect.  In fact, small galaxies (i.e., galaxies with small
scale lengths) will have a greater possibility to be selected in a
sample of edge-on galaxies than in a face-on galaxy sample, due to the
line-of-sight integration through the galaxy discs.  Therefore, his
explanation seems to be valid, since we do not detect any lack of
late-type disc galaxies with small scale lengths in our sample. 

Finally, Fig.  \ref{type.fig}c shows the distribution of absolute {\it
I} and {\it K}-band magnitudes with Hubble type.  Since in edge-on disc
galaxies the luminosity is, at least for the later-type galaxies
($T>2$), dominated by the disc luminosity, and thus the relation between
the disc and the total absolute magnitude is approximately linear, we
could have deduced the distribution of absolute magnitudes from Figs. 
\ref{type.fig}a and b (de Jong 1996b):
\begin{equation}
M_{\rm abs} \approx M_{\rm disc} \propto \mu_0 - 2.5 \log (2 \pi h_R^2),
\end{equation}
where $\mu_0$ is the (edge-on) disc central surface brightness, and
$M_{\rm abs}$ and $M_{\rm disc}$ are the absolute magnitudes of the
total galaxy and the disc component only, respectively. 

Because the scale length distribution of Fig.  \ref{type.fig}b does not
show any dependence on type, the distribution of absolute magnitudes
reflects the distribution of central surface brightnesses with type,
again consistent with de Jong (1996b). 

\subsubsection{Dust}

\begin{figure*}
\hspace*{0.7cm}
\psfig{figure=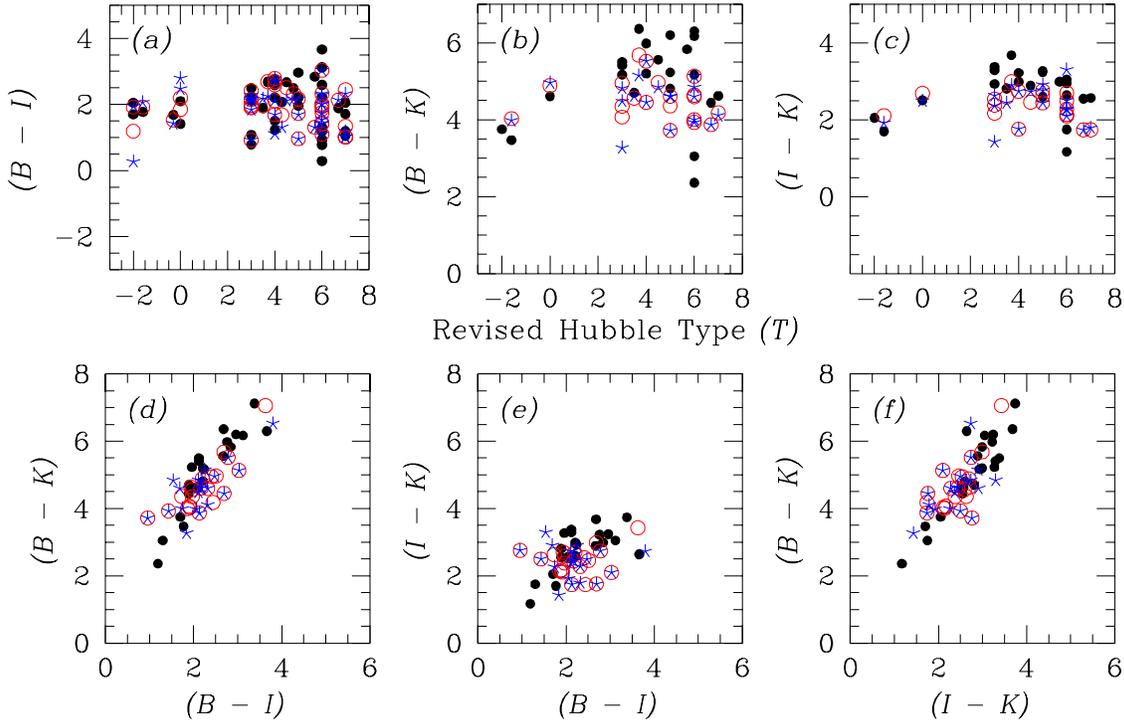,width=16cm}
\vspace*{-5.8cm}
\caption[]{\label{colours.fig}Colour distributions of the sample
galaxies: {\it (a)--(c)} Galaxy colours as a function of revised Hubble
type; {\it (d)--(f)} Colour-colour diagrams. The filled dots denote
central galaxy colours (based on the extrapolated central surface
brightnesses), the open circles represent the galaxies' total colours,
and the stars show the colours of the discs. The galaxy colours have
been corrected for Galactic foreground extinction, based on the {\it
(B--V)} colour excesses predicted by Burstein \& Heiles (1978, 1984).}
\end{figure*}

The distribution of galaxy colours as a function of Hubble type
(corrected for Galactic foreground extinction) is shown in Figs. 
\ref{colours.fig}a--c for the {\it (B--I)$_0$, (B--K)$_0$,} and {\it
(I--K)$_0$} colours, respectively.  We show the central disc colours,
derived from the extrapolated central surface brightnesses, the colours
of the total galaxy and those of the disc, denoted by different symbols. 
It is clear, that these colours follow the same distributions, although
the scatter is large.  In particular for the {\it (B--K)$_0$} and {\it
(I--K)$_0$} colour distributions the colours derived from the total
apparent magnitudes are systematically redder than the other two colour
distributions.  This is caused by dust, because the disc apparent
magnitudes and the central surface brightnesses were obtained from
ellipse fits to the outer disc regions, where the amount of dust is
significantly lower than in the central plane.  De Jong (1996b) notices
a clear correlation between the colours of his 86 face-on spiral
galaxies and galaxy type (but with a large scatter), whereas we do not
find any such correlation.  However, since our sample galaxies are
heavily contaminated by dust, such a correlation, if intrinsic to the
galaxies, may well be hidden by the interstellar dust. 

In Figs.  \ref{colours.fig}d--f we present colour-colour diagrams for
our sample galaxies, for the colours described above.  Although the
distributions follow each other closely and the correlations are
generally tight, the correlation between total galaxy colours is
systematically offset, again due to dust.  This evidence is supported by
the very red {\it (B--K)$_0$} colours of most of our galaxies, which are
redder than those of bright giant ellipticals (with $B-K \sim 4.3$
[Peletier et al.  1994, 1995a]). 

Therefore, Fig.  \ref{colours.fig} is completely dominated by dust
effects; intrinsic colour gradients in the discs of galaxies, due to
either metallicity or age gradients are too small to be observable, see
also Sect.  \ref{radgrads.sect}. 

\section{Discussion}

\subsection{Main observational results}

From the statistical analysis of the global structure parameters of our
sample of edge-on disc galaxies in the {\it B, I,} and {\it K} bands, in
this paper we present the following main observational results:

\begin{itemize}

\item We found a correlation between the $h_R/z_0$ ratio and the revised
Hubble type (Fig.  \ref{ratio_type.fig}a), in the sense that galaxies
become systematically thinner when going from S0's to Sc's, whereas the
later types (later than Sc) seem to be at least as thin as the Sc's. 

\item On average the colour gradients indicated by the scale length
ratios increase from type Sa to at least type Scd.  For galaxy types
later than Scd, the colour gradients seem to become smaller again. 

\item The distribution of (edge-on) {\it K}-band disc central surface
brightnesses is rather flat, although with a large scatter.  However,
the latest-type sample galaxies ($T > 6$) show an indication that their
disc central surface brightnesses may be fainter than those of the
earlier types.  Most likely, this effect is not the result of dust
extinction. 

\item We do not find a clear correlation between galaxy type and
integrated, disc, or central colours.  The scatter in the distributions
is probably caused by dust.

\end{itemize}

\subsection{Kinematic constraints from structure analysis}

As was discussed in Sect.  \ref{ratio.sec}, once the $h_R/z_0$ ratio is
known, one may be able to determine the maximum rotation of a disc from
measurements of the vertical disc dispersion, although the theoretical
predictions are based on assumptions that show large intrinsic scatter
themselves.  Van der Kruit \& Searle (1982a) found a mean scale
parameter ratio of $h_R / z_0 = 5.0 \pm 1.8$ for their sample of 8
edge-on disc galaxies (predominantly later-type spirals) in the optical
{\it J} band, whereas de Grijs \& van der Kruit (1996) reported a mean
ratio of $h_R / z_0 = 5.9 \pm 0.4$ based on the {\it I}-band
characteristic length scales in their sample of 7 edge-on disc galaxies. 
These ratios are larger than the mean ratio found in this paper
($\langle h_R/z_0 \rangle = 3.7 \pm 1.3$). 

To fit the observations of the maximum rotational velocity of
disc-dominated Sc galaxies, $h_R/z_0$ needs to be of order 10 (Bottema
1993).  The discrepancy between the observational results and the
theoretical prediction leads Bottema (1993) to the conclusion that, for
realistic $h_R/z_0$ ratios, the stellar velocity dispersions only allow
the disc to have a (theoretical) maximum rotation of on average $(63 \pm
10)$\% of the observed maximum rotation.  This is significantly less
than implied by results of the so-called maximum disc fits (e.g., van
Albada \& Sancisi 1986) that are widely used to model galaxy rotation
curves, which predict a maximum-disc rotation of 85-90\% of the observed
rotation (Bottema 1993). 

Recent estimates of the scale parameters of our Galaxy (e.g., Kent et
al. 1991; see Sackett [1997] for an overview), yield a scale parameter
ratio of $(h_R/z_0)_{\rm Gal} = 5.3 \pm 0.5$, which is consistent with
the ratios published previously as well as with the data for external
galaxies presented in this paper.  However, although this ratio for our
Galaxy is significantly less than the value of order 10 required
theoretically (Bottema 1993), Sackett (1997) argues that new structural
and kinematical constraints are consistent with a Galactic maximum disc
with reasonable mass-to-light ratio, of $2 \le M/L_V \le 7$. 

In Sect.  \ref{scalepars.sect} we showed that the $h_R/z_0$ ratio of our
sample galaxies correlates with revised Hubble type.  This effect is not
likely to be caused by the presence of a thick disc in our earlier-type
sample galaxies, since we used the thin disc scale parameters to derive
the correlation (see also de Grijs \& Peletier 1997).  Part of the
effect may be due to the distribution of the scale lengths among our
sample galaxies, in the sense that the earlier-type galaxies tend to
have smaller scale lengths.  Moreover, van der Kruit \& Searle (1982a)
argue that the scale height is linearly dependent on the fraction of
stellar to gas mass.  Under this assumption, it is expected that the
vertical scale parameter in the later-type galaxies is smaller than that
for the earlier types.  However, we cannot confirm this assumption on
the basis of our observations. 

\subsection{Radial colour gradients and extinction}
\label{radgrads.sect}

In this paper we show that the mean scale length ratios for the
later-type ($T > 2$) disc galaxies in our sample range from $h_I/h_K =
1.15 \pm 0.19$ to $h_B/h_K = 1.65 \pm 0.41$, indicating large colour
gradients in the discs.  When comparing our results to those published
previously we have to keep in mind the nature of our sample: since it
consists solely of edge-on disc galaxies, the observed scale length
ratios can be expected to be larger than similar ratios obtained from
samples including less highly inclined galaxies. 

Most previously published scale length ratios favour large colour
gradients in galaxy discs:
\begin{itemize}

\item For a large sample of face-on galaxies, Elmegreen \& Elmegreen
(1984) found a mean ratio of $h_B/h_I = 1.16 \pm 0.47$.  Using Evans'
(1994) models this corresponds to an average ratio of {\it B} to {\it
K}-band scale length of about 1.32. 

\item Peletier et al.  (1994, 1995a) present the results of a study of
37 Sb and Sc galaxies (uniform in orientation on the sky), for which
they show that the {\it B} to {\it K}-band scale length ratio varies
between 1.2 and 2, with a mean ratio of $h_B/h_K = 1.49 \pm 0.29$,
comparable to our results. 

\item From de Jong's (1996a) scale length determinations of 86 face-on
spiral galaxies, we find an average {\it B} to {\it K}-band scale length
ratio $h_B/h_K = 1.22 \pm 0.23$

\item From a study of the prototypical dusty galaxy NGC 253, Forbes \&
DePoy (1992) find $h_B/h_H = 1.25 \pm 0.05$.

\end{itemize}
Large colour gradients in galaxy discs between {\it B} and {\it I} band
have also been reported by Kent (1986).  However, from a sample of 33
disc galaxies, van der Kruit (1991) reported a small scale length ratio
between the photographic {\it J} and {\it F} bands, $h_J/h_F = 1.07 \pm
0.13$.

The main problem when comparing scale lengths and scale length ratios
determined by different authors is the radial fitting range used to
derive the scale lengths (Sect.  \ref{hlength.sect}).  However, if scale
length ratios are calculated from scale lengths determined over the same
radial fitting range in each passband, the differences between different
determinations should be $\sim 10$\% at maximum (Peletier et al. 1994). 

Peletier et al.  (1994, 1995a,b) argue that scale length ratios due to
stellar population changes are of order 1.1--1.2 in the blue --
near-infrared range.  Two lines of evidence were used to arrive at this
result: first, from observations of $T < 1$ type galaxies without much
visible dust by Balcells \& Peletier (1994), it was found that $h_B/h_I
= 1.04 \pm 0.05$, corresponding to {\it B} to {\it K}-band scale length
ratios of at most $h_B/h_K = 1.08 \pm 0.10$, using stellar population
models of, e.g., Arimoto \& Yoshii (1986).  From metallicity gradients
from H{\sc ii} regions in galaxy discs, Peletier et al.  (1994, 1995a)
argue that $h_B/h_K$ is likely of order 1.17, using a simple single-age
stellar population model.  In the {\it I} vs.  {\it K} range this
contribution is likely to be less.  Our observations of the scale length
ratios in our earliest-type sample galaxies ($T<1$) support this
evidence (Fig.  \ref{hratios.fig}). 

{\sl Therefore, the observed scale length ratios largely represent the
galaxies' dust content.}

Although the scale length ratios indicate an increasing dust content for
galaxy types later than about $T=3$ (Sb), the data suggests that the
distribution of scale length ratios flattens or even decreases towards
later types, as was also found from optical depth measurements (e.g., de
Grijs et al. 1997).  This result is in accordance with the distribution
found by Peletier \& Balcells (1996). 

\section{Summary and Conclusions}

The main conclusions we can draw from the analysis presented in this
paper are:

\begin{itemize}

\item We found a correlation (although with a large scatter) between the
$h_R/z_0$ ratio and galaxy type, in the sense that galaxies become
systematically thinner when going from S0's to Sc's, whereas the later
types (Sd's) seem to be at least as thin as the Sc's, but likely
thicker. 

\item The average values found for the $h_R/z_0$ ratio are significantly
smaller than the theoretical prediction of about 10 needed for
maximum-disc rotation.

\item On average the colour gradients indicated by the scale length
ratios increase from type Sa to at least type Scd.  For galaxy types
later than Scd, the colour gradients seem to become smaller again. 

\item The observed scale length ratios (and thus the radial colour
gradients) largely represent the galaxies' dust content. 

\item The distribution of (edge-on) {\it K}-band disc central surface
brightnesses is rather flat, although with a large scatter.  However,
the latest-type sample galaxies ($T > 6$) show an indication that their
disc central surface brightnesses may be fainter than those of the
earlier types.  Most likely, this effect is not the result of dust
extinction. 

\item We do not find a clear correlation between galaxy type and
integrated, disc, or central colours.  The scatter in the distributions
is probably caused by dust. 

\end{itemize}

\section*{Acknowledgements}
 
The input of Reynier Peletier, Piet van der Kruit and Roelof Bottema has
been invaluable for the structure and contents of this paper in its
present form. This work is based on observations obtained at the
European Southern Observatory, La Silla, Chile.


\begin{thebibliography}{}

\bibitem[]{} Aoki, T.E., Hiromoto, N., Takami, H., Okamura, S., 1991,
PASJ 43, 755
\bibitem[]{} Arimoto, N., Yoshii, Y., 1986, A\&A 107, 135
\bibitem[]{} Balcells, M., Peletier, R.F., 1994, AJ 107, 135
\bibitem[]{} Barnaby, D., Thronson Jr., H.A., 1992, AJ 103, 41
\bibitem[]{} Barteldrees, A., Dettmar, R.-J., 1994, A\&AS 103, 475
\bibitem[]{} Bottema, R., 1993, A\&A 275, 16
\bibitem[]{} Burstein, D., Heiles, C., 1978, ApJ 225, 40
\bibitem[]{} Burstein, D., Heiles, C., 1984, ApJS 54, 33
\bibitem[]{} Buser, R., 1978, A\&A 62, 411
\bibitem[]{} Byun, Y.I., Freeman, K.C., Kylafis, N.D., 1994, ApJ 432, 114
\bibitem[]{} Carlberg, R.G., 1987, ApJ 322, 59
\bibitem[]{} Carter, B.S., Meadows, V.S., 1995, MNRAS 276, 734
\bibitem[]{} Davies, J.I., 1990, MNRAS 244, 8
\bibitem[]{} de Grijs, R., 1997, Ph.D. Thesis, Groningen University
\bibitem[]{} de Grijs, R., Peletier, R.F., 1997, A\&A 320, L21
\bibitem[]{} de Grijs, R., Peletier, R.F., van der Kruit, P.C., 1997,
A\&A 327, 966
\bibitem[]{} de Grijs, R., van der Kruit, P.C., 1996, A\&AS 117, 19
\bibitem[]{} de Jong, R.S., 1996a, A\&AS 118, 557
\bibitem[]{} de Jong, R.S., 1996b, A\&A 313, 45
\bibitem[]{} de Jong, R.S., 1996c, A\&A 313, 377
\bibitem[]{} de Jong, R.S., van der Kruit, P.C., 1994, A\&AS 106, 451
\bibitem[]{} de Vaucouleurs, G., de Vaucouleurs, A., Corwin, H.G., Jr.,
Buta, R.J., Paturel, G., Fouqu\'e, P., 1991, Third Reference Catalogue of
Bright Galaxies, Springer-Verlag: New York {\bf (RC3)}
\bibitem[]{} Disney, M.J., Davies, J.I., Phillipps, S., 1989, MNRAS 239,
939
\bibitem[]{} Elmegreen, D.M., Elmegreen, B.G., 1984, ApJS 54,127
\bibitem[]{} Evans, R., 1994, MNRAS 266, 511
\bibitem[]{} Fall, S.M., 1983, in: Internal Kinematics and Dynamics of
Galaxies, IAU Symposium 100, ed.  Athanassoula, E., Dordrecht: Reidel,
p.  391
\bibitem[]{} Fall, S.M., Efstathiou, G., 1980, MNRAS 193, 189
\bibitem[]{} Forbes, D.A., DePoy, D.L., 1992, A\&A 259, 97
\bibitem[]{} Freeman, K.C., 1970. ApJ 160, 811
\bibitem[]{} Giovanelli, R., Haynes, M.P., Salzer, J.J., Wegner, G., Da
Costa, L.N.,  Freudling, W., 1994, AJ 107, 2036
\bibitem[]{} Guthrie, B.N.G., 1992, A\&AS 93, 255
\bibitem[]{} Hamabe, M., Kodaira, K., Okamura, S., Takase, B., 1979, 
PASJ 31, 431
\bibitem[]{} Hamabe, M., Kodaira, K., Okamura, S., Takase, B., 1980, 
PASJ 32, 197
\bibitem[]{} Hamabe, M., Wakamatsu, K., 1989, ApJ 339, 783   
\bibitem[]{} Hegyi, D.J., Gerber, G., 1979, in: Photometry, Kinematics,
and Dynamics of Galaxies, ed. Evans, D.S., Austin: Univ. of Texas
Astron. Dept., p. 119
\bibitem[]{} Hubble, E., 1926, ApJ 64, 321
\bibitem[]{} Huizinga, J.E., 1994, Ph.D. Thesis, Groningen University
\bibitem[]{} Jensen, E.B., Thuan, T.X., 1982, ApJS 50, 421 
\bibitem[]{} Kent, S.M., 1986, AJ 91, 1301
\bibitem[]{} Kent, S.M., Dame, T.M., Fazio, G., 1991, ApJ 378, 131
\bibitem[]{} Knapen, J.H., van der Kruit, P.C., 1991, A\&A 248, 57
\bibitem[]{} Kuchinski, L.E., Terndrup, D.M., 1996, AJ 111, 1073
\bibitem[]{} Kylafis, N.D., Bahcall, J.N., 1987, ApJ 317, 637
\bibitem[]{} Landolt, A.U., 1992, AJ 104, 340
\bibitem[]{} Larson, R.B., Tinsley, B.M., 1978, ApJ 219, 46
\bibitem[]{} Lauberts A., Valentijn, E.A., 1989, The Surface Photometry
Catalogue of the ESO-Uppsala Galaxies, ESO {\bf (ESO-LV)}
\bibitem[]{} Mathewson, D.S., Ford, V.L., Buchhorn, M., 1992, ApJS 81, 413
\bibitem[]{} Mathewson, D.S., Ford, V.L., 1996, ApJS 107, 97 (AAS CD-ROM
Series, Vol. 7)
\bibitem[]{} Peletier, R.F., 1993, A\&A 271, 51
\bibitem[]{} Peletier, R.F., Balcells, M., 1996, in: Spiral Galaxies in
the Near-IR, eds. Minnitti, D., Rix, H.-W., ESO/MPA Workshop, p. 48
\bibitem[]{} Peletier, R.F., Balcells, M., 1997, New Astr. 1, 349
\bibitem[]{} Peletier, R.F., Valentijn, E.A., Moorwood, A.F.M.,
Freudling, W., 1994, A\&AS 108, 621
\bibitem[]{} Peletier, R.F., Valentijn, E.A., Moorwood, A.F.M.,
Freudling, W., 1995a, in: The Opacity of Spiral Disks, ed. Davies, J., 
NATO Advanced Research Workshop, p. 243
\bibitem[]{} Peletier, R.F., Valentijn, E.A., Moorwood, A.F.M.,
Freudling, W., Knapen, J.H., Beckman, J.E., 1995b, A\&A 300, L1
\bibitem[]{} Peletier, R.F., Willner, S.P., 1992, AJ 103, 1761
\bibitem[]{} Richter, O.-G., Tammann, G.A., Huchtmeier, W.K., 1987, A\&A
171, 33
\bibitem[]{} Rieke, G.H., Lebofsky, M.J., 1985, ApJ 288, 618
\bibitem[]{} Sackett, P.D., 1997, ApJ 483, 103
\bibitem[]{} Sandage, A., 1986, A\&A 161, 89
\bibitem[]{} Sasaki, T., 1987, PASJ 39, 849 
\bibitem[]{} Schmidt, K.-H., Boller, T., 1992, Astron. Nachr. 313, 189
\bibitem[]{} Searle, L., Sargent, W.L.W., Bagnuolo, W.G., 1973, ApJ 179, 427
\bibitem[]{} Shaw, M.A., Gilmore, G., 1990, MNRAS 242, 59
\bibitem[]{} Terndrup, D.M., Davies, R.L., Frogel, J.A., DePoy, D.L., 
Wells, L.A., 1994, ApJ 432, 518
\bibitem[]{} Thuan, T.X., Gunn, J.E., 1976, PASP 88, 543
\bibitem[]{} Tinsley, B.M., Larson, R.B., 1978, ApJ 221, 554
\bibitem[]{} Toomre, A., 1964, ApJ 139, 1217
\bibitem[]{} van Albada, T.S., Sancisi, R., 1986, Philos. Trans. R. Soc.
London, Ser. A 320, 447
\bibitem[]{} van der Kruit, P.C., 1987, A\&A 173, 59
\bibitem[]{} van der Kruit, P.C., 1991, in: The World of Galaxies, eds. 
Corwin, H.G., Bottinelli, L., New York: Springer, p. 256
\bibitem[]{} van der Kruit, P.C., Searle, L., 1981a, A\&A 95, 105 
\bibitem[]{} van der Kruit, P.C., Searle, L., 1981b, A\&A 95, 116
\bibitem[]{} van der Kruit, P.C., Searle, L., 1982a, A\&A 110, 61
\bibitem[]{} van der Kruit, P.C., Searle, L., 1982b, A\&A 110, 79
\bibitem[]{} Wainscoat, R.J., Freeman, K.C., Hyland, A.R., 1989, ApJ 337, 163
\bibitem[]{} Wainscoat, R.J., Hyland, A.R., Freeman, K.C., 1990, ApJ 348, 85
\bibitem[]{} Wainscoat, R.J., Cowie, L.L., 1992, AJ 103, 332

\end{thebibliography}
\end{document}